\documentclass[]{rsproca_new}
\usepackage{graphicx}
\usepackage{amsmath} 
\usepackage{amssymb} 
\usepackage{mathrsfs} 
\usepackage{bm}
\usepackage{float}
\usepackage[numbers,sort&compress]{natbib}

\newcommand{\mrm}[1]{\mathrm{#1}}
\newcommand{\bbm}[1]{\boldsymbol{\mrm{#1}}}
\newcommand{\ex}[0]{ \hat{\bbm{e}}_x }
\newcommand{\ey}[0]{ \hat{\bbm{e}}_y }
\newcommand{\et}[0]{ \hat{\bbm{e}}_\theta }
\newcommand{\er}[0]{ \hat{\bbm{e}}_r }

\newcommand{\un}[0]{ \hat{\bbm{n}}}
\newcommand{\nby}[0]{\mathrm{N}_{\mathrm{B}y}}
\newcommand{\nbx}[0]{\mathrm{N}_{\mathrm{B}x}}

\newcommand{\p}[0]{\partial }

\graphicspath{{./figures/}}

\begin{document}

\title{Long-wave equation for a confined ferrofluid interface: Periodic interfacial waves as dissipative solitons}

\author{Zongxin Yu and Ivan C.\ Christov}

\address{School of Mechanical Engineering, Purdue University, West Lafayette, UN 47907, USA}

\subject{applied mathematics, fluid mechanics, wave motion}


\corres{Ivan C. Christov\\
\email{christov@purdue.edu}}

\begin{abstract}
We study the dynamics of a ferrofluid thin film confined in a Hele-Shaw cell, and subjected to a tilted nonuniform magnetic field. It is shown that the interface between the ferrofluid and an inviscid outer fluid (air) supports traveling waves, governed by a novel modified Kuramoto--Sivashinsky-type equation derived under the long-wave approximation. The balance between energy production and dissipation in this long-wave equations allows for the existence of dissipative solitons. These permanent traveling waves' propagation velocity and profile shape are shown to be tunable via the external magnetic field. A multiple-scale analysis is performed to obtain the correction to the linear prediction of the propagation velocity, and to reveal how the nonlinearity arrests the linear instability. The traveling periodic interfacial waves discovered are identified as fixed points in an energy phase plane. It is shown that transitions between states (wave profiles) occur. These transitions are explained via the spectral stability of the traveling waves. Interestingly, multiperiodic waves, which are a non-integrable analog of the double cnoidal wave, are also found to propagate under the model long-wave equation. These multiperiodic solutions are investigated numerically, and they are found to be long-lived transients, but ultimately abruptly transition to one of the stable periodic states identified.
\end{abstract}


\begin{fmtext}

\section{Introduction}
Immiscible fluid flows confined in Hele-Shaw cells have been investigated extensively during the past several decades \cite{MMM19}. Going back to the classical work by Saffman and Taylor \cite{Saffman1958}, interest has focused on the dynamics of the sharp interface between the fluids \cite{BKLST86}. The interface's displacement, when the motion of the fluids is normal to

\end{fmtext}


\maketitle

\noindent
the unperturbed interface, has been of particular interest to most studies, specifically viscous fingering instabilities and finger growth \cite{Homsy1987}. By contrast, Hele-Shaw flows in which the main flow direction is \emph{parallel} to the fluid interface has received less attention. Early work by Zeybek and Yortsos~\cite{ZY91,ZY92} considered such a parallel flow in a horizontal Hele-Shaw cell, both theoretically and experimentally. They found that, in the limit of large capillary number and under the long-wave assumption, interfacial waves between the two viscous fluids in this setup are governed by a set of coupled Korteweg--de Vries (KdV) and Airy equations. Similarly, Charru and Fabre~\cite{CF93} investigated periodic interfacial waves between two viscous fluid layers in a Couette flow, in which case the long-wave equation was found to be of Kuramoto--Sivashinsky (KS) type. Subsequently, experimental work by Gondret and co-workers~\cite{GR97,MGRR03} demonstrated traveling waves in a parallel flow in a vertical Hele-Shaw cell. In this case, the phenomenon is well-described by a modified Darcy equation accounting for inertial effects, in which context a Kelvin--Helmholtz instability for inviscid fluids was found~\cite{PH02,HP05}. These prior studies considered fluids that are not responsive to external stimuli.

Ferrofluids (also known as ``magnetic fluids'' \cite{S74,R87}), on the other hand, are colloidal suspensions of nanometer-sized magnetic particles dispersed in a nonmagnetic carrier fluid. These fluids are typically Newtonian but respond to applied magnetic fields, which is of particular interest in the present work. The linear theory of the Kelvin--Helmholtz instability for unconfined ferrofluids was developed by Rosensweig~\cite{R13_Ferrohydrodynamics}, which revealed how the strength of the applied magnetic field (on top of the velocity difference and viscosity contrast between the fluid) enters the threshold for instability. Miranda and Widom~\cite{MW00} extended this result to a parallel ferrofluid flow in a vertical Hele-Shaw cells under an external non-tilted magnetic field and deduced that the magnetic field does not affect the propagation speed of waves. Using a perturbative weakly nonlinear analysis, Lira and Miranda~\cite{LM12} further extended the latter analysis by adopting an in-plane tilted applied magnetic field, showing that the wave speed speed is sensitive to the angle. Such a field was shown to generate nonlinear traveling surface waves between a ferrofluid and an inviscid fluid (such as air). Jackson and Miranda~\cite{JM07} introduced a ``crossed'' magnetic field (with perpendicular and azimuthal components) to influence the mode selection for a ferrofluid drop confined in a horizontal Hele-Shaw cell. Beyond Hele-Shaw configurations, Seric \textit{et al.}~\cite{SAK14} derived a long-wave equation to model dewetting of a two-dimensional thin film resulting from the interaction between a uniform applied magnetic field and disjoining pressure. More recently, Yu and Christov \cite{YC21} conducted fully nonlinear simulations, using a vortex sheet Lagrangian method, of ferrofluid droplets in a horizontal Hele-Shaw cell. They showed that nonlinear periodic waves can be generated on the ferrofluid interface by tuning an external magnetic field's orientation. In their analysis, the nonlinear wave propagation speed was well predicted by perturbation theory, showing that the magnetic field can set the wave speed and induce rotation of the droplet. Despite the recent work and interest on how tilted magnetic fields generate nonlinear waves on ferrofluid interfaces, a model long-wave equation, to describe these phenomena is still lacking. Such a reduced-order (``low-dimensional'') model would provide deeper insight into the nonlinear wave dynamics and the mechanisms that sustain them \cite{KT07}.

\begin{figure}
 \centering
 \includegraphics[keepaspectratio=true,width=\columnwidth]{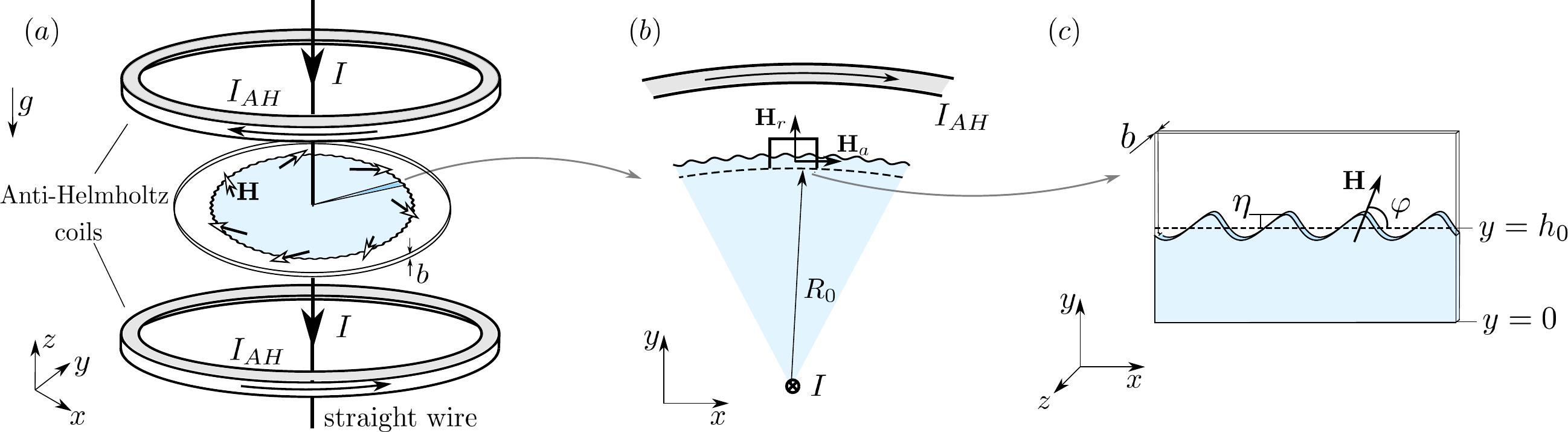}
 \caption{(a) Schematic of a Hele-Shaw cell (width $b$) confining a thin ferrofluid film, with unperturbed depth $h_0$. An azimuthal magnetic field $\bbm H_{a}$ can be produced by a long wire conveying an electric current $I$, adapted from \cite{YC21}. A radial magnetic field $\bbm H_{r}$ can be produced by a pair of anti-Helmholtz coils with equal and apposite currents $I_{AH}$. (b) Top view of Hele-Shaw cell, with boxed region to be studied as a periodic domain. (c) In the local Cartesian coordinates, the external magnetic field $\bbm H$ is tilted at angle $\varphi$ with respect to the $x$-axis, and it acts to deforms the interface at $y=f(x,t)=h_0+\eta(x,t)$. The fluid exterior to the thin film (\textit{e.g.}, air) is assumed to have negligible viscosity and velocity. {Gravity acts in the $-z$-direction, but its effects are negligible for the in-plane interface evolution}.}
 \label{fig:config}
\end{figure}

To this end, in this work, our goal is to derive a novel \emph{model} long-wave equation \cite{H74,AH76,ODB97} to describe the nonlinear wave dynamics on a confined ferrofluid interface. First, we reexamine the problem proposed in \cite{YC21} by considering a thin ferrofluid film in Cartesian coordinates (as shown in Fig.~\ref{fig:config}), subjected to an in-plane tilted magnetic field, which makes an arbitrary angle with the unperturbed (flat, horizontal) interface. Next, a perturbation analysis similar to that for shallow water waves, valid for small wave amplitudes and long wavelengths, is conducted. We show that the interfacial waves are governed by a modified equation of the Kuramoto--Sivashinsky (KS) type. Although the KS equation is usually mentioned in the context of the work by Kuramoto and Tsuzuki \cite{KT76}, on phase turbulence in reaction-diffusion systems, and the work by Sivashinsky~\cite{S77}, on wrinkled flame front propagation, the equation was first derived by Homsy~\cite{H74} for thin liquid films (see also the discussions in \cite{CV95,KKP15}). On the other hand, the generalized KS equation, which additionally contains a dispersion term, has been derived in the context of a wide variety of falling thin film problems~\cite{KRSV12}, for which the driving force is typically gravity. In the present work, the novel feature is the non-invasive forcing of the ferrofluid by a magnetic field, which leads to a new type of generalized KS equation that captures the myriad of nonlinear effects (interestingly, in the absence of the Hopf-like convective nonlinearity found in the traditional KS-type equations) on long-wave evolution in one-dimension (1D).

This paper is organized as follows. {\S}\ref{sec:governing eq} introduces the governing equations of the parallel ferrofluid thin film flow confined to a horizontal Hele-Shaw cell. In {\S}\ref{sec:long wave eq}, the long-wave equation for the interface dynamics is derived, exposing the key parameters governing the physics. In {\S}\ref{sec:linear, energy budget}, linear and weakly nonlinear analyses are conducted to understand the wave dynamics. An energy budget for the nonlinear traveling wave solution of the long-wave equation is obtained, showing that a dissipative soliton can propagate under the novel balance of surface tension and magnetic forces in this system. The effects of the key parameters on the wave profile and its propagation are discussed in {\S}\ref{sec:wave profile}. Then, {\S}\ref{sec:state transition} considers the transition between different nonlinear states, and their spectral stability, to address the pattern selection problem. Additionally, propagating multi-periodic waves are uncovered numerically, and their persistence is investigated. Finally, conclusions and avenues for further work are summarized in {\S}\ref{sec:conclusion}.


\section{Mathematical model and governing equations}
\label{sec:governing eq}

Building on our previous work \cite{YC21}, we study the dynamics of interfacial waves on a thin ferrofluid film, confined in the transverse direction within a Hele-Shaw cell with gap thickness $b$, as shown in Fig.~\ref{fig:config}. In the reference configuration, the unperturbed interface is at $r=R_0+h_0$. The entire cell is subjected to a radially-varying external magnetic field via a long wire carrying an electric current $I$ through the origin. This current produces an azimuthal magnetic field component $\bbm H_a=\frac{I}{2\pi} \frac{1}{r} \et$. Then, anti-Helmholtz coils can be used to produce a radial magnetic field component $\bbm H_r=\frac{H_0}{R_0}r \er$, where $H_0$ is strength of the magnetic field at $r=R_0$ \cite{LM16,ALM18,YC21}. Now assume that $R_0 \gg h_0$, where $h_0$ is a characteristic `depth' of the ferrofluid film at rest. Under this ``small film curvature" assumption~\cite{RBG21}, the nonuniform magnetic field $\bbm H = \bbm H_a + \bbm H_r$ can be approximated in locally Cartesian coordinates as:
\begin{equation}
    \bbm H \simeq \frac{I}{2\pi}\frac{1}{(R_0+y)}\ex + \frac{H_0}{R_0}(R_0+y) \ey.
    \label{eq:H_cartesian}
\end{equation}

From Eq.~\eqref{eq:H_cartesian}, we understand that a magnetic body force $\propto |\bbm M| \bbm\nabla |\bbm H|$ acts on the thin film, where $\bbm M$ is the magnetization vector of the ferrofluid. For the purposes of studying the interface and shape dynamics \cite{MO04,RE06,LM16,ACLM19}, we assume that the ferrofluid is uniformly magnetized, and the magnetization is colinear with the external field, \textit{i.e.}, $\bbm M=\chi \bbm H$, where $\chi$ is the constant magnetic susceptibility. Since the applied field is spatially varying, \textit{i.e.}, $\bbm\nabla|\bbm H| \ne \bbm 0$, then $\bbm\nabla|\bbm H|$ becomes the main contribution to the magnetic body force. According to the prior literature, this observation leads us to neglect the effect of the demagnetizing field in comparison.

It is straightforward to show by standard methods (see, \textit{e.g.}, \cite{LM16} and the references therein) that neglecting inertial hydrodynamic terms, enforcing the no-slip condition on the confining boundaries (transverse to the flow) of the Hele-Shaw cell, and averaging across the gap (\textit{i.e.}, over $z$) yields a modified ``Darcy's law'' that governs this flow~\cite{LM16}:
\begin{equation}
\bar{\bbm v} = -\frac{b^2}{12\mu_f} \bbm\nabla\left(p-\Psi\right), \qquad
\bbm\nabla \cdot \bar{\bbm v} = 0,\qquad -\infty<x<\infty,\quad 0\le y \le f(x,t).
\label{eq:governing}
\end{equation}
Here, $p$ is the hydrodynamic pressure in the film, $\mu_f$ is the ferrofluid's dynamic viscosity, $\Psi=\mu_0 \chi |\bbm H|^2/2$ is a scalar potential accounting for the magnetic body force (such that $p-\Psi$ is a modified pressure), and $\mu_0$ is the free-space permeability. {Gravity acts in the $-z$-direction, but it is neglected due to the narrow confinement.} Both fluids are considered incompressible. The viscosity of the ``upper'' fluid is considered negligible (\textit{i.e.}, it is considered inviscid, as would be the case with air), so the flow outside the ferrofluid film is not considered. We denote by $\bar{\bbm v} = u(x,y,t) \ex + v(x,y,t) \ey$ the $z$-averaged velocity field in the ``lower'' fluid (the ferrofluid).

At the interface, having neglected the dynamics of the upper fluid, the pressure is given by a modified Young--Laplace law~\cite{R13_Ferrohydrodynamics,BCM10}:
\begin{equation}
p=\sigma \kappa-\frac{\mu_0}{2}(\bbm M \cdot \un)^2 
\quad \text{on} \quad  y=f(x,t),
\label{eq:pressure jump}
\end{equation}
where $\sigma$ is the constant surface tension, and $\kappa \equiv -{f_{xx}}/{(1+f_x^2)^{3/2}}$ is the curvature of the surface $y=f(x,t)$ ($x$ and $t$ subscripts denote partial derivatives). The second term on the right-hand side of Eq.~\eqref{eq:pressure jump} is the magnetic normal traction~\cite{R13_Ferrohydrodynamics,BCM10}, where $\un = ({-f_x},{1})/{\sqrt{1+f_x^2}}$ denotes the upward unit normal vector to the interface. This contribution, due to the projection of $\bbm M$ onto $\un$, induces unequal normal stress on either side of the profile's peaks on the perturbed interface, thus breaking the initial equilibrium and leading to wave propagation \cite{YC21}.

A kinematic boundary condition is also imposed at the interface:
\begin{equation}
    v = f_t + u f_x \quad \text{on} \quad  y=f(x,t),
    \label{eq:kinematic_bc}
\end{equation} 
which requires that the film boundary is a material surface. 
The no-penetration condition 
\begin{equation}
v =0 \quad \text{on} \quad  y=0
\label{eq:bottom_bc}
\end{equation}
is imposed at the ``bottom'' of the layer, which is the material surface at $r=R_0$ in the original radial coordinates (Fig.~\ref{fig:config}), that maps to $y=0$.

Introducing the potential $\phi=p-\Psi-\Psi_0$, where the constant
\begin{equation}
\Psi_0=-\frac{\mu_0}{2}\chi \frac{H_0^2}{R_0^2}(R_0+h_0)^2(1+\chi)-\frac{\mu_0}{2}\chi \frac{I^2}{4\pi^2}\frac{1}{(R_0+h_0)^2}    
\end{equation}
accommodates the trivial solution, and combining the two equations in \eqref{eq:governing} together, the governing equation becomes Laplace's equation:
\begin{equation}
\nabla^2\phi=0.
\label{eq:Laplace}
\end{equation}
From Eqs.~\eqref{eq:pressure jump} and \eqref{eq:kinematic_bc}, Eq.~\eqref{eq:Laplace} is subject to the following boundary conditions on $y=f(x,t)$:
\begin{subequations}\begin{align}
&v = f_t+uf_x,\\
&\phi + 
\frac{\mu_0 \chi}{2}\frac{H_0^2(R_0+y)^2 }{R_0^2}
+\frac{\mu_0 \chi}{2}\frac{I^2}{4\pi^2(R_0+y)^2}
+\Psi_0 \\
&\qquad\qquad = \sigma  \kappa - \frac{\mu_0 \chi^2}{2} \left[
		\frac{I^2}{4\pi^2(R_0+y)^2} \frac{f_x^2}{1+f_x^2}
		+\frac{H_0^2(R_0+y)^2}{R_0^2} \frac{1}{1+f_x^2}
		-\frac{IH_0}{\pi R_0} \frac{f_x}{1+f_x^2}
		\right]. \nonumber
\end{align}\label{eq:dyn_BC}\end{subequations}
\vspace{-4mm}


\section{Derivation of the long-wave equation}\label{sec:long wave eq}
\subsection{Expansion of the potential and non-dimensionalization}
To reduce the governing equations to a single partial differential equation (PDE) for the surface deformation $\eta$, we expand $\phi$ in a power series in $y$, a standard approach for small amplitude surface deformations (see, \textit{e.g.}, \cite{LR12}):
\begin{equation}
\phi(x,y,t)=\sum_{n=0}^\infty y^n\phi_n(x,t).
\label{eq:phi_expand}
\end{equation}
Substituting this expansion into Laplace's equation~\eqref{eq:Laplace} generates a recursion relation $\phi_{n,xx}+(n+2)(n+1)\phi_{n+2}=0$.
On the other hand, since 
$\phi_y=\sum_{n=1}^\infty ny^{n-1}\phi_n(x,t)$,
the constraint at the bottom (\textit{i.e.}, Eq.~\eqref{eq:bottom_bc}) requires that $\phi_1=0$, which eliminates the odd terms from the expansion. Hence, we can simplify Eq.~\eqref{eq:phi_expand} as:
\begin{equation}
\phi(x,y,t)=\sum_{m=0}^\infty \frac{(-1)^my^{2m}}{(2m)!}g^{(2m)}(x,t), \qquad 
g^{(2m)}(x,t)\equiv\frac{\p^{2m}}{\p x^{2m}}\phi_0(x,t).
\label{eq:phi_expand2}
\end{equation}

Let $a$  be the typical amplitude scale for the surface deformation $\eta(x,t)$. Now, we introduce
the following non-dimensionalization:
\begin{equation}
\begin{alignedat}{5}
x&\mapsto\ell x , \quad
&y &\mapsto h_0y ,\quad
&t &\mapsto \frac{12\mu_f \ell^3}{\sigma b^2} t,\quad
&\eta &\mapsto a\eta,\quad\\
u&\mapsto\left(\frac{a}{h_0} \right)\frac{\sigma b^2}{12\mu_f\ell^2}u,\quad
&v &\mapsto\left(\frac{a}{h_0} \right)\left(\frac{h_0}{\ell} \right)\frac{\sigma b^2}{12\mu_f\ell^2}v,\quad
&\phi &\mapsto \left(\frac{a}{h_0} \right)\frac{\sigma}{\ell} \phi,\quad
&g &\mapsto \left(\frac{a}{h_0} \right)\frac{\sigma}{\ell} g,
\end{alignedat}
\end{equation}
where $\ell$ is the horizontal length scale.
Next, we define the small parameters of the model
\begin{equation}
\delta := \frac{h_0}{\ell},\qquad
\epsilon := \frac{a}{h_0},\qquad
\varepsilon := \frac{h_0}{R_0},
\label{eq:small_params}
\end{equation}
corresponding to a wavelength parameter, an amplitude parameter, and a magnetic field gradient parameter, respectively. To implement the upcoming asymptotic expansion, a long wavelength $\delta \ll 1$ and small amplitude $\epsilon \ll 1$ approximation is made \cite{B66}. (Although it is possible to also derive arbitrary-amplitude long-wave equations \cite{B66,AH76}, Homsy \cite{H74} argued that the distinguished limit of $\epsilon\ll1$ leads to the \emph{model} equations capturing the essential physics.) Note that $\varepsilon \ll 1$ is determined by the geometric configuration; specifically, $R_0$ is chosen sufficiently large to allow the Cartesian approximation, but small enough to ensure that $\bbm\nabla|\bbm H|$ is still the dominant term in the magnetic body force \cite{MO04,RE06,LM16}. Note that demagnetization can still be neglected because it can be made arbitrarily small via the thickness $b$ \cite{ACLM19}.

The scaled potential obeys:
\begin{equation}
\phi_{xx}+ \delta^{-2}\phi_{yy} = 0,\qquad
u=-\phi_x, \qquad
v=-\delta^{-2}\phi_y,\qquad
\label{eq:uvphi relation}
\end{equation}
and, to $\mathcal{O}(\delta^2)$, the scaled and truncated Eq.~\eqref{eq:phi_expand2} yields:
\begin{equation}
 \phi = g-\frac{1}{2}\delta^2y^2 g_{xx}, \qquad
u = -g_x+\frac{1}{2}\delta^2 y^2 g_{xxx}, \qquad
v = yg_{xx}-\frac{1}{6}\delta^2 y^3 g_{xxxx},
\label{eq:uvphi tranc}
\end{equation}
consistent with Eq.~\eqref{eq:uvphi relation}.

\subsection{Boundary conditions and reduction of the governing equations}
{Under the above assumptions on the small parameters, the leading-order terms involve $\epsilon$, $\varepsilon^2$, and $\delta^2$. Without making further assumption on their relative scalings (thus, keeping cross-terms as well), the corresponding kinematic and dynamic boundary conditions \eqref{eq:dyn_BC} on the fluid--fluid interface become:}
\begin{subequations}\label{eq:BC_simp}\begin{alignat}{2}
v&=\eta_t+\epsilon u \eta_x &&\quad \text{on}\quad y=1+\epsilon \eta(x,t), \\
\phi &= B_1 \eta -\delta \eta _{xx}+\delta  B_2\eta_x+\epsilon \delta^2 B_3 \eta_x^2
-B_4\epsilon\eta^2
&&\quad \text{on}\quad y=1+\epsilon \eta(x,t),
\label{eq:boundary with ABCD}
\end{alignat}\end{subequations}
where $B_n$ are constants (see electronic supplementary material {\S}A for their expressions). Importantly, the constants are functions of the  magnetic Bond numbers:
\begin{equation}
\nbx =\frac{\mu_0 \chi}{2}\frac{I^2}{4\pi^2}\frac{1}{R_0^2}\frac{\ell}{\sigma},\qquad
\nby =\frac{\mu_0 \chi}{2}H_0^2\frac{\ell}{\sigma},
\end{equation}
which quantify the ratios of the magnitudes of the $x$ and $y$ components of the magnetic body force to the surface tension force.

Before proceeding further in the analysis, we rewrite the boundary conditions from Eqs.~\eqref{eq:BC_simp} to hold at $y=1$ through Taylor series expansions of $u$, $v$, and $\phi$:
\begin{subequations}\label{eq:perturbed1}\begin{alignat}{2}
v+v_y\epsilon \eta &=\eta_t+\epsilon (u+u_y\epsilon \eta)\eta_x &&\quad \text{on}\quad y=1,\\
\phi+\phi_y\epsilon \eta&= B_1 \eta -\delta \eta_{xx} +\delta B_2 \eta_x +\epsilon \delta^2 B_3\eta_x^2
-B_4\epsilon\eta^2
&&\quad \text{on}\quad y=1.
\end{alignat}\end{subequations}
With the relations in Eq.~\eqref{eq:uvphi relation}, Eqs.~\eqref{eq:perturbed1} can be rewritten, within the assumed order, as 
\begin{subequations}\label{eq:bc v phi}\begin{alignat}{2}
v &= \eta_t - \epsilon \left\{ [B_1 \eta_x+ \delta(B_2 \eta_{xx} - \eta_{xxx} )]\eta \right\}_x &&\quad \text{on}\quad y=1,\\
\phi &= B_1 \eta + \delta(B_2 \eta_{x}  - \eta_{xx} )-B_4\epsilon\eta^2 +\epsilon \delta^2 (B_3 \eta _x^2 +\eta \eta_t) &&\quad \text{on}\quad y=1.
\end{alignat}\end{subequations}

Combining Eqs.~\eqref{eq:uvphi tranc}, evaluated at $y=1$, and Eqs.~\eqref{eq:bc v phi} allows us to eliminate $g(x,t)$, and the dynamics of the interface $\eta(x,t)$ is governed by
\begin{equation}\begin{aligned}
\eta_t&= (-\varepsilon\alpha-\varepsilon^2\vartheta) \eta_{xx} +\delta (\beta \eta_{xxx} - \eta_{xxxx} )
+\epsilon  \left\lbrace[(-\varepsilon\alpha-\varepsilon^2\vartheta) \eta_x+\delta(\beta \eta_{xx}  -\eta _{xxx} )]\eta\right\rbrace_x\\
&\phantom{=}
-\epsilon \frac{1}{2} \varepsilon^2\vartheta (\eta^2)_{xx}
+\frac{1}{2}\delta^2\left[ \eta_{xxt} -\frac{1}{3}(-\varepsilon\alpha-\varepsilon^2\vartheta) \eta_{xxxx} \right]\\ 
&\phantom{=}+\epsilon \delta^2 \left[ (\gamma \eta_x^2 +\eta \eta_t)_{xx} -\frac{1}{4}(-\varepsilon\alpha-\varepsilon^2\vartheta) (\eta^2)_{xxxx} +\frac{1}{12}\varepsilon^2\vartheta (\eta^2)_{xx} \right],
\end{aligned}\label{eq: general eta}
\end{equation}
where
\begin{subequations}\begin{align}
\alpha &= 2[\nby(1+\chi) - \nbx ],\label{eq:alpha}\\
\beta  &= 2\chi \sqrt{\nbx \nby},\\
\gamma &= \chi (\nby - \nbx),\\
\vartheta &= 2[(1+\chi)\nby + 3\nbx],
\end{align}\label{eq: main parameters}\end{subequations}
are now the governing dimensionless parameters of the model, beyond the previously defined small quantities in Eq.~\eqref{eq:small_params}. Note that Eq.~\eqref{eq: general eta} is a general expression of the interface dynamics without any assumption about the relation between the (three) small parameters $\epsilon$, $\delta$, and $\varepsilon$. To obtain a \emph{model} equation, in sense of \cite{H74}, we must consider the relevant distinguished limit.

\subsection{The model long-wave equation}
Next, we seek to simplify the governing  Eq.~\eqref{eq: general eta} in the distinguished asymptotic limit(s) of interest. For $\varepsilon=\mathcal{O}(\delta^2)$, without loss of generality, we let $\varepsilon=\delta^2$ and conduct another rescaling:
\begin{equation}
\eta \mapsto \eta/\epsilon, \qquad t \mapsto t/\delta
\label{eq:long time scal}
\end{equation}
to describe the long-time evolution (as expected, since we focus on traveling wave solutions). From Eq.~\eqref{eq: general eta}, the interface evolution equation for $\varepsilon=\mathcal{O}(\delta^2)$ can be written as:
\begin{equation}
\eta_t = -\delta\alpha \eta_{xx} + \beta \eta_{xxx} - \eta_{xxxx} 
+[(-\delta\alpha \eta_x+\beta \eta_{xx}  - \eta_{xxx} )\eta]_x + \delta(\gamma \eta_x^2 )_{xx}.
\label{eq:e=delta2}
\end{equation}

The long-wave equation for $\varepsilon=\mathcal{O}(\delta)$ has similar structure as Eq.~\eqref{eq:e=delta2} (see electronic supplementary material {\S}B for details), so that in this study we will focus on Eq.~\eqref{eq:e=delta2}, which is a modified generalized KS equation. The main difference lies in the dispersion and nonlinear terms. Whereas the KS equation features the Hopf nonliterary $\eta\eta_x$ (as do the KdV and Burgers equations), Eq.~\eqref{eq:e=delta2} does not. Instead, the last two terms on the right-hand side of Eq.~\eqref{eq:e=delta2} depict a more complicated nonlinearity introduced \emph{almost entirely} by the magnetic forces. One of the latter terms, $\propto (\eta_x\eta)_x$, is similar to the term due to the Maragoni effect in the so-called Korteweg--de Vries--Kuramoto--Sivashinsky--Velarde equation \cite[Eq.~(6)]{CV95}.
We note in passing that this term, together with the term $\propto (\eta_{xxx}\eta)_x$ and $\eta\eta_x$, also appear in the nonlinear terms of the model equation for interfacial periodic waves in \cite[Eq.~(9)]{CF93}. As in present study, the  $(\eta_{xxx}\eta)_x$ nonlinearity arises from surface tension. However, while $(\eta_x\eta)_x$ in \cite[Eq.~(9)]{CF93} comes about from inertia, in our model equation this term arises from magnetic forces. Meanwhile, the role of the linear terms is well known, as in KS: $\eta_{xx}$ is responsible for the instability at large scales, while $\eta_{xxxx}$ provides dissipation at small scales. As in the generalized KS equation, the KdV-like term $\eta_{xxx}$ in Eq.~\eqref{eq:e=delta2} leads to dispersion.


\section{Stability of the flat state and nonlinear energy budget}
\label{sec:linear, energy budget}

\subsection{Linear growth rate and weakly nonlinear mode coupling}
Let $\eta(x,t)=\sum_{k=-\infty}^\infty \eta_k(t) e^{ikx}$ be the Fourier decomposition of the surface elevation on the periodic domain  $x\in[0,2\pi]$. 
Then, substituting the Fourier series into Eq.~\eqref{eq:e=delta2}, we immediately obtain:
\begin{equation}
\dot{\eta_k}=\Lambda (k)\eta_k +\sum_{k'}F(k,k') \eta_{k'}\eta_{k-k'} ,
\label{eq:Fourier EOM}
\end{equation}
where the overdot denotes a time derivative, $k\neq 0$, $k'\neq 0$, $\eta_{k=0}=0$, $i=\sqrt{-1}$, and
\begin{subequations}\begin{align}
\Lambda(k)&=\delta\alpha k^2 -k^4 -i\beta k^3,
\label{eq:growth rate}\\
F(k,k')&= \delta\alpha kk'-i \beta kk'^2-kk'^3
+2\delta \gamma (k^2k'^2 -kk'^3).
\label{eq:nonlinear growth rate}
\end{align}\label{eq:lambda_F}\end{subequations}
Recalling the definition of $\alpha$ from Eq.~\eqref{eq:alpha}, the real part of the linear growth rate $\Re[\Lambda(k)]$ indicates that the $y$-component of the magnetic field  $\propto (1+\chi)\nby$ is destabilizing, while the $x$-component $\propto\nbx$ and surface tension are stabilizing. Weakly-nonlinear mode coupling at the second-order is accounted for by the function $F$. Note that the terms in $\Re[\Lambda(k)]$ from Eq.~\eqref{eq:growth rate} above are quite similar to the ones in \cite{YC21} (for a radial geometry), apart from being multiplied by an additional power of $k$.

\begin{figure}
    \centering
    \includegraphics[keepaspectratio=true,width=\columnwidth]{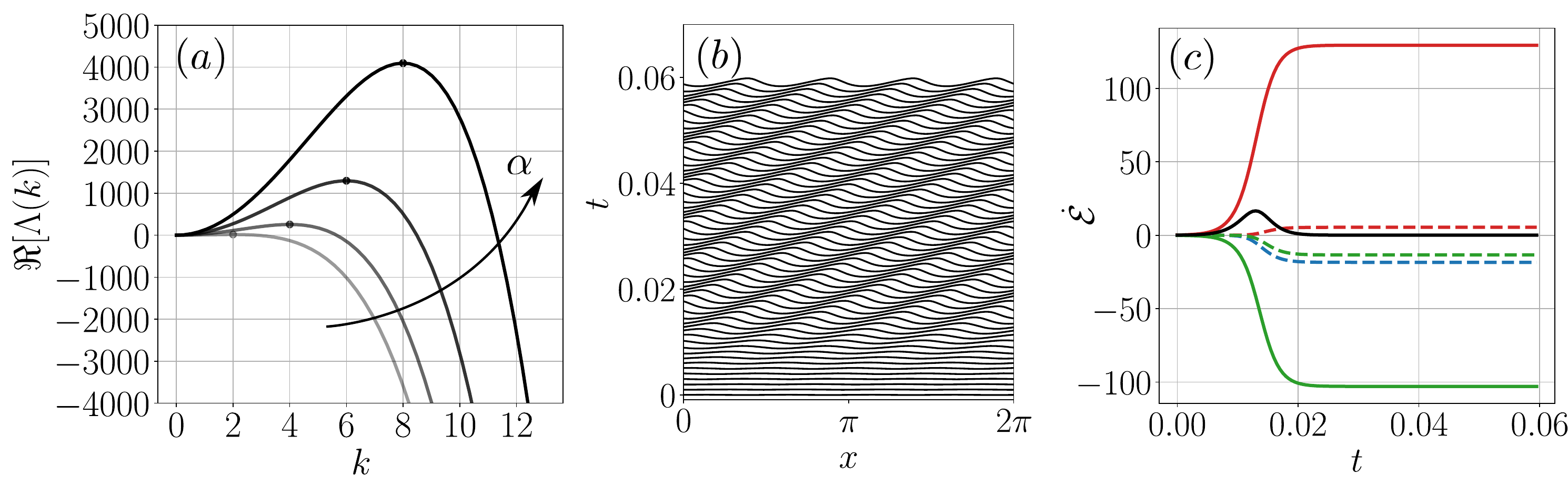}
    \caption{(a) Real part of the linear growth rate $\Re[\Lambda(k)]$ as a function of the wavenumber $k$ for $\delta \alpha=8$, $32$, $72$ and $128$; the markers denote the most unstable mode $k_m$. (b) The nonlinear evolution of the interface from a small perturbation of the flat base state [$\eta(x,0)=0.01\sin (4x)$] into a permanent traveling wave with $\delta \alpha=32$, $\beta=16$, and $\gamma$ determined by Eq.~\eqref{eq: main parameters} accordingly. (c) Energy budget of the nonlinear traveling generation process shown in (b); the red curve represents the $\delta\alpha$ term, the green curve represents the surface tension term, and the blue curve represents the $\beta$ term from the PDE~\eqref{eq:e=delta2}. The contribution of the linear term is denoted by the solid curves while the dashed curves represent the nonlinear term(s). The black curve in (c) shows the sum of these components, which is seen to approach zero as the wave evolves into a dissipative soliton.} 
    \label{fig:energy budget}
\end{figure}

The most unstable mode $k_m$ satisfies:
\begin{equation}
\left. \frac{d \Re[\Lambda(k)]}{dk}\right|_{k=k_m}=0 \qquad\Longleftrightarrow\qquad 2k_m^2=\delta\alpha,
\label{eq:km}
\end{equation}
which implies the important role of $\delta\alpha$ on stability. Figure~\ref{fig:energy budget}(a) shows examples of how $\delta\alpha$ controls the most unstable mode and determines the range of linearly unstable modes (for which $\Re[\Lambda(k)]>0$). We will show that $k_m$ (and $\delta \alpha$) can be used to predict the possible states (period of the nonlinear interfacial wave), and it is helpful for selecting suitable initial conditions that evolve into (nonlinear) traveling wave solutions.

The imaginary part of the linear growth $\Im[\Lambda(k)]$ rate reveals the phase velocity of each mode:
\begin{equation}
    v_p(k)=-\Im[\Lambda(k)]/k=\beta k^2.
    \label{eq:phase vel}
\end{equation}
Perturbations to the flat base state of the film can propagate with velocity controlled by the coupling term $\beta=\chi\sqrt{\nbx\nby}$ (and, since $v_p=v_p(k)$, they also experience dispersion). Here, $\beta$ results from the magnetic normal stress due to the asymmetric projection of the $x$- and $y$-components of the magnetic force onto the interface. Changing the direction of the $x$-component of $\bbm{H}$ will reverse the sign of these terms, \textit{i.e.}, $\beta\mapsto-\beta$. The linear analysis indicates that such wavepackets will either decay or blow-up exponentially according to the sign of $\Re[\Lambda(k)]$. However, below we will show, through simulations of the governing PDE, that this linear instability is arrested by nonlinearity.

\subsection{Nonlinear energy balance and the dissipative soliton concept}
\label{sec:energy balance}

The energy method \cite{St04} can be applied to any PDE to understand the stability of its solutions. For example, the energy method was used to establish stability and uniqueness of generic ferrofluid flows \cite{VK02}. Here, we employ this approach to understand the stability of the traveling wave in our model long-wave equation, which features both damping and gain. Multiplying Eq.~\eqref{eq:e=delta2} by $\eta$, and integrating by parts over $x\in[0,2\pi]$, yields an energy balance:
\begin{equation}
\dot{\mathcal{E}}
=\int_0^{2\pi} 
 \delta\alpha \eta_x^2
- \eta_{xx}^2
+\delta\alpha \eta_{x}^2\eta
+\frac{1}{2}\beta\eta_{x}^3 
-\eta \eta_{xx}^2
\,dx,
\label{eq:energy budget}
\end{equation}
where $\mathcal{E}(t) \equiv \frac{1}{2}\int_0^{2\pi}\eta(x,t)^2 \,dx$ denotes the total energy of the wave field. The $\delta\alpha\eta_x^2$ term on the right-hand side of Eq.~\eqref{eq:energy budget} produces energy, while the surface tension term $-\eta_{xx}^2$ acts as a sink. This result matches well with the observation regarding the linear growth rate, \textit{i.e.}, that the destabilizing $\delta\alpha$ term is balanced by the (stabilizing) surface tension ($\delta\alpha k^2>0$ and $-k^4<0$ in Eq.~\eqref{eq:growth rate}).
The linear dispersion term conserves energy and thus drops out of Eq.~\eqref{eq:energy budget}. Meanwhile the sign of the three remaining terms is indeterminate \textit{a priori}. Figure~\ref{fig:energy budget}(c) show the evolution of the various terms on the right-hand side of Eq.~\eqref{eq:energy budget} for the solution $\eta(x,t)$ shown in Fig.~\ref{fig:energy budget}(b).

Eventually, all curves in Fig.~\ref{fig:energy budget}(c) become independent of time. In general, we expect that, for some distinguished solutions $\eta(x,t)$, $\dot{\mathcal{E}}=0$ holds exactly. If this is the case for one of the traveling wave solutions, then they are classified as \emph{dissipative solitons} in the sense of \cite{CV95}. Dissipative solitons are expected to be long-lived stable structures. We wish to address if such structures arise in our model of a ferrofluid interface subjected to a magnetic field.

From the energy analysis in Eq.~\eqref{eq:energy budget}, we can conclude that $\alpha$, $\beta$, and $\gamma$ are the three key parameters controlling the wave propagation and existence of the dissipative soliton. Recall that these three parameters, which show up in Eq.~\eqref{eq:e=delta2}, are given as combinations of the physical parameters (\textit{i.e.}, $\chi$, $\nbx$, $\nby$), as per Eq.~\eqref{eq: main parameters}. In particular, $\alpha$ and $\beta$ in the linear terms of Eq.~\eqref{eq:e=delta2} are expected to strongly affect the stability and the characteristics of the traveling wave profile. We explore this issue next through numerical simulations.

\subsection{Numerical simulation strategy for the governing long-wave PDE}
\label{sec:numerical_method}
 
To understand the nonlinear interfacial wave dynamics, in the upcoming sections below, we solve Eq.~\eqref{eq:e=delta2} numerically using the pseudospectral method \cite{STW11}. For the linear terms, the spatial derivatives are evaluated using the fast Fourier transform (FFT) with $N=512$, while the nonlinear terms are inverted back to the physical domain (via the inverse FFT), evaluated, and then transformed back to Fourier space. The modified exponential time-differencing fourth-order Runge--Kutta (ETDRK4) scheme \cite{KT05}, which is stable and accurate for stiff systems \cite{STW11}, is adopted for the time advancement. Grid and time-step convergence of the numerical scheme was established (see electronic supplementary material {\S}C). Figure~\ref{fig:energy budget}(b) shows an example evolution from the infinitesimal perturbation of the flat state, to the formation of a nonlinear traveling wave.


\section{Nonlinear periodic interfacial waves: propagation velocity and shape}
\label{sec:wave profile}

As discussed in {\S}4\ref{sec:energy balance}, $\alpha$ and $\beta$ play an important role in the energy balance. In this section, we investigate their effects on the traveling wave's propagation velocity and the wave profile (shape). Before we start, it is helpful to discuss the physical meaning of these parameters, which can be useful in designing control strategies in practice. 

First, as explained above, $\beta$ is the coupling term resulting from the asymmetry of the surface force on the perturbed interface. This parameter is also closely related to the orientation of the magnetic field. To understand this point better, let
\begin{equation}
\rho=\nbx+\nby,\qquad q=\nbx/\rho.
\label{eq:q rho}
\end{equation}
Here, $\rho\propto|\bbm H|^2$ relates to the magnitude of the magnetic field at $R_0$, and $q=\cos^2\varphi$, where $\varphi$ is the angle of $\bbm H$ with respect to the flat interface (recall Fig.~\ref{fig:config}). With $\chi=1$, the main parameters can be rewritten as:
\begin{equation}
 \alpha = 2\rho(2-3q),\qquad \beta = 2\rho\sqrt{q(1-q)},\qquad \gamma =\rho(1-2q).
 \label{eq:parameters rho q }
\end{equation}

In this study, we restrict ourselves to magnetic fields with small $x$-component magnitude, with $q\in[0,0.06]$, \textit{i.e.}, $\varphi\in[0.42\pi,{\pi}/{2}]$. For this choice, $\alpha\approx4\rho$, $\beta\approx 2\rho \sqrt{q}$, and $\gamma \approx \rho$. Hence, controlling $\alpha$ is equivalent to controlling the magnitude of the magnetic field, while $\beta$ is sensitive to the orientation. Note that two independent variables will set the dynamics, and in this section we will control $\alpha$ and $\beta$, with $\gamma$ determined by Eq.~\eqref{eq:parameters rho q }. 
Furthermore, in the numerical studies below, we will use one initial condition, $\eta(x,t=0)=0.01\sin(k_0x)$, with an initial perturbation wavenumber $k_0=4$, and we will only consider $\delta=0.1$. These perturbations will first grow, then become arrested by saturating nonlinearity \cite{BP98}, and finally lead to a permanent traveling wave. The latter is of interest in this section.

\subsection{Propagation velocity}
\subsubsection{Linear prediction and nonlinear expression}
Lira and Miranda \cite{LM12} reported that the propagation velocity of interfacial ferrofluid waves in a Cartesian configuration in a vertical Hele-Shaw cell is sensitive to the magnetic field's angle. The fully nonlinear simulations of a radial configuration in a horizontal Hele-Shaw cell in \cite{YC21} further showed that this velocity can be well predicted by the linear phase velocity, which is determined by the coupling term of azimuthal and radial magnetic field components in that work. In this study, we examine how this coupling term, which is captured by our parameter $\beta$ (and closely related to the angle $\varphi$ of the magnetic field), controls the nonlinear wave propagation velocity.

A permanent traveling wave profile takes the form $\eta(x,t)=\Theta(k x-\omega t)$, where $v_f=\omega/k$ is its propagation (phase) velocity. The modes' complex amplitudes can be expressed as $\eta_{k}(t)=c_{k}e^{-i\omega(k) t}$, with constant $c_{k}\in\mathbb{C}$ accounting for their relative phases. A nonlinear traveling wave profile would consist of a fundamental mode $k_f$ and its harmonics $nk_f$ ($n\in\mathbb{Z}^+$), with $\omega(nk_f)=n\omega(k_f)$, so that the phase velocity can be evaluated as $v_p(nk_f,t)=n\omega(k_f)/nk_f = v_f$. The average $v_p$ of the first five harmonics is used to calculate $v_f^N$ for the nonlinear simulation.
Meanwhile, the linear phase velocity $v_f^L=v_p$ is given by Eq.~\eqref{eq:phase vel}. 

Figure~\ref{fig:beta assk}(a) shows the comparison of the nonlinear propagation velocity and the linear prediction for $\delta\alpha=32$. The fundamental mode, computed as $k_f=4$ from the simulation, sets the linear propagation velocity as $v_f^L=\beta k^2=16\beta$. It is surprising to see that the actual nonlinear propagation velocity can be well fit by the straight line $v_f^N=14.05\beta$ with small variance $\sigma^2=0.005$, even if the wave profiles changes with $\beta$ dramatically, as shown in Fig.~\ref{fig:beta assk}(c). This curious correction is not as trivial as it looks, as the nonlinear phase velocity can be evaluated \textit{a posteriori} through Eqs.~\eqref{eq:lambda_F} as:
\begin{equation}
    v_f^N = \beta
    \left\lbrace 
    k^2
    +\sum_{k'}k'^2\Re\left[\frac{\eta_{k'}\eta_{k-k'}}{\eta_{k}}\right]
    -\frac{k}{\beta}\sum_{k'}[\delta\alpha k'-k'^3+2\delta\gamma(kk'^2-k'^3)]\Im\left[\frac{\eta_{k'}\eta_{k-k'}}{\eta_{k}}\right]
    \right\rbrace,
\label{eq:nonlinear phase velocity}
\end{equation}
{where $v_f^N=\Im[\mathcal{N}(k,k')/k]$ is derived from the propagator operator  $\mathcal{N}(k,k')=\lambda(k)+\sum_{k'}F(k,k')\eta_{k'}\eta_{k-k'}/\eta_{k}$ derived in Eq.~\eqref{eq:Fourier EOM}, such that $\dot{\eta}_k=\mathcal{N}(k,k') \eta_{k}$. The terms arising from the summation over $k'$ represent the nonlinear effects.}
When a traveling wave solution is obtained, $\eta_{k'}\eta_{k-k'}/\eta_{k}=c_{k'}c_{k-k'}/c_{k}$ becomes independent of time, and the nonlinear phase velocity can be evaluated from Eq.~\eqref{eq:nonlinear phase velocity}, knowing $c_k$ from the propagation profile's Fourier decomposition. (This is equivalent to our approach in electronic supplementary material {\S}D. That approach is simpler, therefore the results hereafter follow the approach from electronic supplementary material {\S}D for simplicity and clarity.) 

The correction in Eq.~\eqref{eq:nonlinear phase velocity} is an \textit{a posteriori} result, and it is accurate but not obvious how it changes the pre-factor $k^2=16$ into $14.05$. Nevertheless, the strong, linear correlation between $v_f^N$ and $\beta$ for the chosen parameters of interest is the key point. 

\subsubsection{Multiple-scale analysis and velocity correction}
To better understand the linear correlation between $v_f^N$ and $\beta$, an analytical approximation can be obtained via a multiple-scale analysis of the harmonic wave \cite{KC96}. However, when subject to the current parameters (\textit{i.e.}, $k_m=4$ as the most unstable mode), the linear instability poses difficulties when using a standard travailing wave ansatz. We introduce the critical wave number $k_c$ so that $\Re[\Lambda(k_c)]=0$ $\Rightarrow$ $k_c^2=\delta\alpha$. The linear theory predicts that all $k<k_c$ are unstable. Thus, we assume that the $\delta \alpha$ in the linear term is slightly larger than $k_f^2$, thereby making $k_f=4$ marginally unstable, and also the unique unstable mode. In other words:
\begin{equation}
    \delta\alpha=k_f^2+\mathfrak{e}^2 \varkappa,
\end{equation}
where $\mathfrak{e}\ll 1$ is small perturbation parameter and $\varkappa>0$ is independent of $\mathfrak{e}$.
We first scale  Eq.~\eqref{eq:e=delta2} to a weakly nonlinear problem by introducing $\eta=\mathfrak{e} \mathcal{Y}$:
\begin{equation}
\mathcal{Y}_t + \left(k_f^2+\mathfrak{e} ^2\varkappa\right) \mathcal{Y}_{xx} - \beta \mathcal{Y}_{xxx} + \mathcal{Y}_{xxxx} =
\mathfrak{e}\left\lbrace 
[(-\delta\alpha \mathcal{Y}_x
+\beta \mathcal{Y}_{xx}  
- \mathcal{Y}_{xxx} )\mathcal{Y}]_x
+ (\delta\gamma \mathcal{Y}_x^2 )_{xx}\right\rbrace.
\label{eq:weak nonlinear}
\end{equation}
Next, we introduce the traveling wave coordinate $\xi=kx-\omega_p t$ of a harmonic wave, where $\omega_p=k^3\beta$ by the linear dispersion relation. We assume that $\mathcal{Y}$ has a multiple-scale expansion of the form
\begin{equation}
\mathcal{Y}=\mathcal{Y}_0(\xi,t_2)+\mathfrak{e}\mathcal{Y}_1(\xi,t_2) + \mathfrak{e}^2\mathcal{Y}_2(\xi,t_2) + \mathcal{O}(\mathfrak{e}^3),
\label{eq: y expansion}
\end{equation}
where the slow time is $t_2=\mathfrak{e}^2 t$. By eliminating the secular term at $\mathcal{O}(\mathfrak{e}^2)$ (see electronic supplementary material {\S}E for details), we obtain the leading-order solution
\begin{equation}
\mathcal{Y}_0=2\mathfrak{a}\cos \left(kx-\omega_pt-k^2 \frac{\Im[Q]}{\Re[Q]}\mathfrak{e}^2 \varkappa t+\mathfrak{b}_0\right),
\label{eq:y_0 sol}
\end{equation}
which gives the phase velocity with the multiple-scales correction as
\begin{equation}
v_f^{MS}=\beta k^2 +\frac{\Im[Q]}{\Re[Q]}k\mathfrak{e}^2 \varkappa,
\label{eq:vf ms}
\end{equation}
where $\mathfrak{a}=\sqrt{\varkappa k^2/\Re{[Q]}}$ is the equilibrium amplitude, $\mathfrak{b}_0$ is an integration constant, and 
\begin{equation}
    Q=\frac{[\delta\alpha -i\beta k+(2\delta\gamma-1)k^2]}
{6k^2+3i\beta k} [-\delta\alpha k^2+5i\beta k^3+(7+4\delta\gamma)k^4].
\end{equation}

\begin{figure}
 \centering
 \includegraphics[keepaspectratio=true,width=0.99\columnwidth]{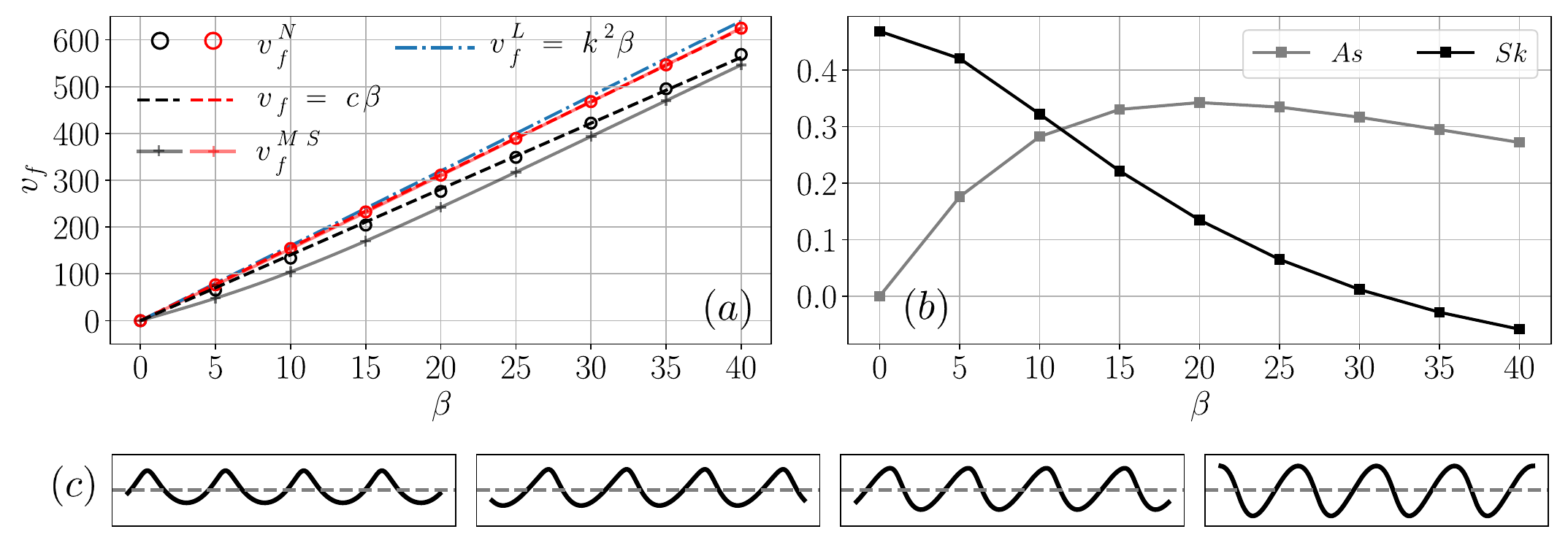}
 \caption{The dependence of (a) the propagation velocity, (b) the asymmetry and skewness on $\beta$ for $\delta\alpha=32$ and $\gamma$ determined by Eq.~\eqref{eq:parameters rho q }, (c) example traveling wave profiles for $\beta=0$, $10$, $20$, $40$ for $\delta\alpha=32$. In (a), the ``$\circ$" denotes the nonlinear velocity evaluated from the simulation; the dashed line fits the nonlinear velocity with slope ``$c$", the solid line marked with ``+" is the velocity prediction from the multiple-scales expansion. Black sets represent results with $\delta\alpha=32$, while red ones are from $\delta\alpha=18$. }
 \label{fig:beta assk}
\end{figure}

Equation~\eqref{eq:vf ms} predicts the propagation velocity of the traveling wave solution when $k_f=4$ is subjected to weak linear instability. The weak linear instability is important to emphasize in this multiple-scales derivation because we assumed $\mathfrak{e}^2 \varkappa \ll 1$.
However, the results of this analysis appear to hold even for stronger linear instability.
As $\delta \alpha$ in the original formulation increases from $k_f^2=16$, $\mathfrak{e}^2 \varkappa$ increases correspondingly. In Fig.~\ref{fig:beta assk}(a), we show two cases with $\delta \alpha=18$ ($\mathfrak{e}^2 \varkappa=2$) and $\delta \alpha=32$ ($\mathfrak{e}^2 \varkappa=16$). For the weakly linearly unstable case ($\delta \alpha=18$), the propagation velocity $v_f^{MS}$ predicted by the multi-scale expansion matches well with the nonlinear velocity $v_f^{N}$, with error less than $0.5\%$. For stronger linear instability, \textit{i.e.}, $\delta \alpha=32$, $v_f^{MS}$ is still qualitatively corrected, but the error is now less than $26\%$ for smaller $\beta$, while the agreement improves for larger $\beta$, with the error reducing to about $5\%$. 

Another message obtained from Fig.~\ref{fig:beta assk} is that, even if the nonlinear propagation velocity $v_f^N$ shows a linear correlation with $\beta$, it is not necessarily linearly related to $\beta$ due to the nonlinearities of the PDE. However, this observation will not change the fact that such linear correlation enables both $v_f^L$ and $v_f^{MS}$ to be good predictors for the wave dynamics (and their possible control via the imposed magnetic field). In this respect, another reason that $v_f^L$ is a good quantitative prediction is the lack of ``inertia'' in this system. In the classical model equations, such as the KS, KdV, and Burgers, the $\eta\eta_x$ term accounts for nonlinear advection, and thus the initial ``mass'' ($\int^{2\pi}_0 \eta\,dx$) sets the velocity. In our system, the nonlinear terms have a similar effect, while the initial ``mass'' $\int^{2\pi}_0 \eta\,dx=0$ due to the definition of $\eta$ as a periodic perturbation. Thus, the propagation velocity is well predicted directly by the dispersion parameter $\beta$.

\subsection{Traveling wave profile}

In addition to setting the propagation velocity, the coupling term $\beta$ (as the source of asymmetry of the magnetic traction force) also strongly affects the shape of the traveling wave. To explore the shape change, as in \cite{YC21}, we introduce the skewness $Sk$ and asymmetry $As$:
\begin{equation}
    Sk(t) =\frac{ \langle\eta(x,t)^3\rangle}{\langle\eta(x,t)^2\rangle^{3/2}},\qquad
    As(t) =\frac{\langle \mathcal{H}[\eta(x,t)]^3\rangle}{\langle\eta(x,t)^2\rangle^{3/2}},
\end{equation}
where $\left\langle\, \cdot\, \right\rangle=\frac{1}{2\pi} \int_0^{2\pi} (\,\cdot\,)\, dx$, and $\mathcal{H}[\,\cdot\,]$ is the Hilbert transform.
$Sk(t)$ quantifies the vertical asymmetry of nonlinear surface water waves~\cite{KCKD00,M13}, about the unperturbed interface, with $Sk>0$ corresponding to narrow crests and flat troughs (and \textit{vice versa} for $Sk<0$). Meanwhile, $As(t)$ quantifies the fore-aft asymmetry of a wave profile \cite{KCKD00,M13}, with $As>0$ corresponding to waves that ``tilt forward'' (in the direction of propagation). 

Figure~\ref{fig:beta assk}(b) shows the effect of $\beta$ on the wave profile for $k_m=4$ (\textit{i.e.}, for $\delta\alpha=32$). The asymmetry is a concave function of $\beta$ with a maximum around $\beta\approx20$. This implies the existence of an ``optimum'' magnetic field angle that allows tuning of the wave profile shape. For the parameters used in Fig.~\ref{fig:beta assk}(c), \textit{i.e.}, $\delta\alpha=32$ and $\beta=0$, $10$, $20$, $40$, correspondingly we have $q=0,3.9\times10^{-3}$, $0.015$, $0.056$, spanning two orders of magnitude of the magnetic field angle parameter. For $\beta=0$ ($q=0$), the profile is symmetric, and we observe that even a small angle of the magnetic field, breaks the fore-aft asymmetry of the  wave profile. The skewness, on the other hand, monotonically decreases with the angle, becoming negative beyond $\beta\approx 30$.

\begin{figure}
 \centering
 \includegraphics[keepaspectratio=true,width=\columnwidth]{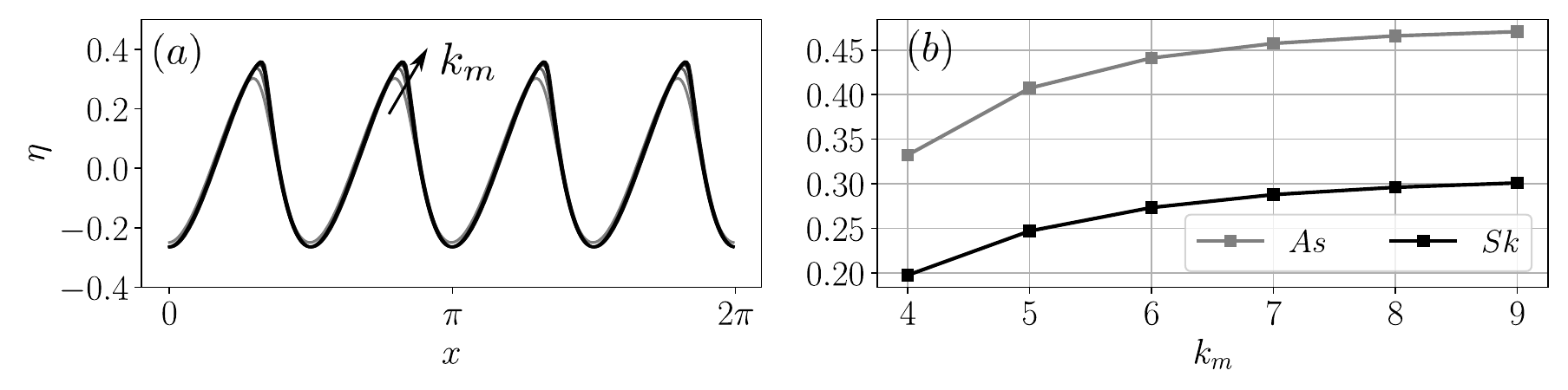}
 \caption{(a) The propagating wave profile with $k_m=4,5,\hdots,9$, $q=0.01$, with the initial perturbation wavenumber $k_0=4$, and $\beta,\gamma$ determined by Eq.~\eqref{eq: main parameters}. (b) The corresponding asymmetry and skewness.}
 \label{fig:alpha assk}
\end{figure}

Figure~\ref{fig:alpha assk} shows how $k_m$ (or, equivalently, $\alpha$ since $\delta$ is fixed) affects the wave profile. When the initial perturbation wavenumber $k_0=4$ is close to the most unstable mode $k_m$, the traveling wave profile maintains the same period as the initial condition. For larger $k_m$, the wave profile exhibits a sharper peak. This sharpening was also observed in \cite[Fig.~3(b,c)]{YC21}, wherein the skewness increases with $k_m$, and the profiles saturate for large values of $k_m$. In \cite{YC21}, the possibly unstable evolution for $k_m$ was not discussed, while the wave studied therein shows ``wave breaking" for large values of the dispersion parameter. Therefore, an open problem that can be addressed with the present long-wave model is the ``asymptotic" behavior of the steepening wave profile ($As$ and $Sk$) with $k_m$. This leads us to a new question: is the range of $k_m$ that allows such period-four waves bounded, or will the shape eventually become unstable (and/or ``break'')? Or, we can reframe the question as: given $k_m$, which states (period of the traveling wave) exist in this system? What about their stability? In the next section, we perform numerical investigations to shed light on these questions.

\section{State transition and stability of traveling waves}\label{sec:state transition}

The KS equation is well known for the chaotic behavior of its solutions. It has been thoroughly investigated within the scope of instability and bifurcation theory \cite{PS91,BJNRZ13,CDS10,KKP15}, yielding a wealth of results on how different dynamical states can be ``reached'' from given initial data, and the transition between such states. Specifically, as the ratio of coefficients of the second- and fourth-order derivative terms (\textit{i.e.}, their relative importance, quantified by $\delta \alpha$ in our model~\eqref{eq:e=delta2}) increases, the KS equation's steady profile exhibits more complexity and dynamical possibilities, and finally the dynamics becomes chaotic. This feature can be understood intuitively from Fig.~\ref{fig:energy budget}(a), wherein higher $\delta \alpha$ allows a wider unstable band for the long waves in the system.

Considering some of the similarities between our long-wave equation~\eqref{eq:e=delta2} and the generalized KS equation, a thorough examination of all the parametric dependencies of the wave (including chaotic) dynamics is not of interest herein. Instead, we focus on showing that the dissipative solitons emerging from perturbations in the linearly unstable band are fixed points in an energy phase plane. Then, we analyze the state transitions via this phase plane, and explain the stability of fixed points via the spectral stability of the wave profiles. {It is noteworthy that, this state transition process is a feature of systems having multi-mode wave solutions. The multi-mode transition process is a generalization of the dynamics studied in the last section, which focused on single-mode evolution.} Finally, we highlight multiperiodic profiles analogous to ``double cnoidal waves'' of the KdV equation.

\subsection{Fixed points in the energy phase plane}
\label{sec:fp_epp}

To reduce the parameter space exploration, in this section we fix $q=0.01$, with $q$ defined in Eq.~\eqref{eq:q rho}, and focus on the dynamics for different $k_m$ only, by controlling the magnitude $\rho=5,20,45,80$. Recall that (as discussed at the beginning of {\S}\ref{sec:wave profile}) there are two independent physical dimensionless groups (\textit{i.e.}, $\nbx$ and $\nby$), so that fixing $q$ and $\rho$ determines all other parameters (\textit{i.e.}, $\alpha$, which sets $k_m$, $\beta$, and $\gamma$). In this subsection, the energy phase plane $(\mathcal{E},\dot{\mathcal{E}})$ (see, \textit{e.g.}, \cite{KKP15}) will be used to identify the traveling wave solutions, which emerge as fixed points with finite $\mathcal{E}$ and $\dot{\mathcal{E}}=0$ (\textit{i.e.}, they are dissipative solitons).

\begin{figure}
 \centering
 \includegraphics[keepaspectratio=true,width=0.9\columnwidth]{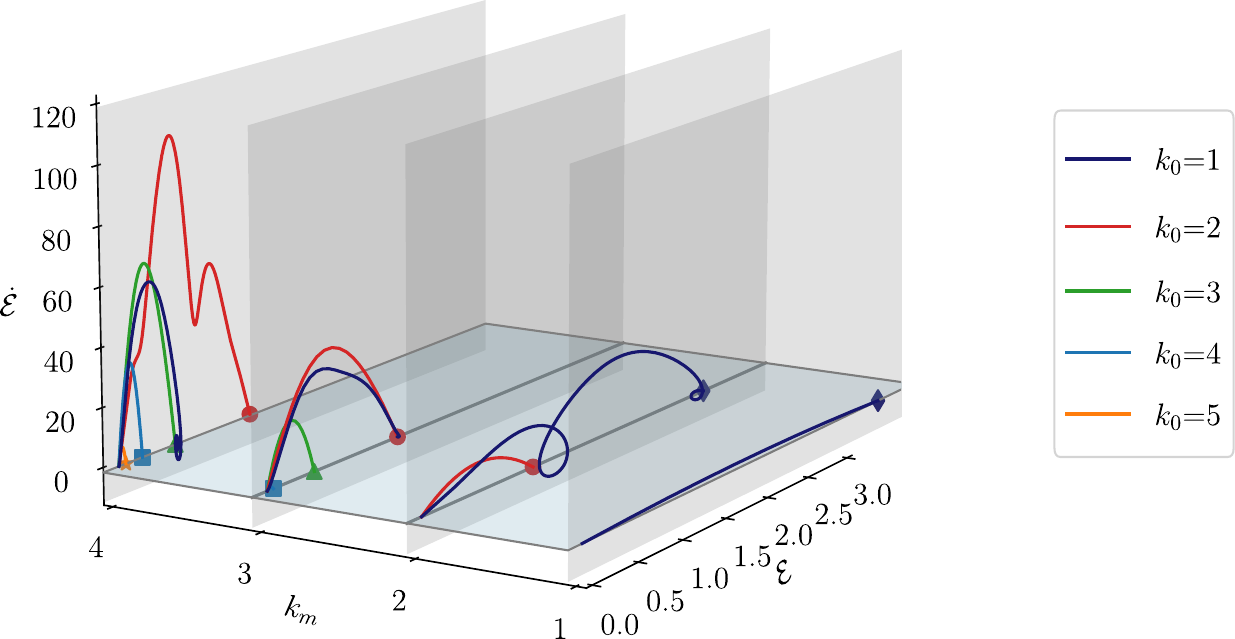}
 \caption{The energy phase plane $(\mathcal{E},\dot{\mathcal{E}})$, showing dynamics in slices corresponding to most unstable wavenumbers $k_m\approx1,2,3,4$ $(\rho=5,20,45,80)$ and, within each slice, the trajectories emerging from initial perturbations with wavenumbers $k_0=1,2,3,4,5$.}
 \label{fig:energy plane 3D}
\end{figure}

The wavenumber range, $k\in(0,k_c]$, of linearly unstable modes can be obtained by solving $\Re[\Lambda(k_c)]=0$ to obtain $k_c=\sqrt{2}k_m$. Figure~\ref{fig:energy plane 3D} shows four slices of the energy phase plane at $k_m=1,2,3,4$. The corresponding maximal linearly unstable modes have wavenumbers $\lfloor k_c\rfloor=1,2,4,5$. As before, the initial condition is selected as a small-amplitude single-mode perturbation: $\eta(x,t=0)=0.01\sin(k_0x)$ with $k_0=1,2,3,4,5$ as the initial wavenumbers. From the nonlinear growth rate in Eq.~\eqref{eq:nonlinear growth rate}, we know that a nonlinear interaction exists only between harmonic modes $nk_0$ ($n\in\mathbb{Z}$) when initializing with the single mode $k_0$. These interacting modes will grow or decay and finally become balanced  harmonic components of the permanent traveling wave profile that emerges.
 
The fundamental mode $k_f$ contains the highest energy, $\eta_{k_f}\eta_{k_f}^*$, in the system. Physically, the wave will exhibit a period-$k_f$ profile, or a ``$k_f$-state.'' For a period-$k_f$ traveling profile, only the harmonic modes $nk_f$ exist in the system. Therefore, the fixed points identified in Fig.~\ref{fig:energy plane 3D} are of different periods for a given $k_m$. In Fig.~\ref{fig:energy plane 3D}, $k_0\neq k_f$ when $k_0=1$, $k_m=3,4$.

For $k_m=1$, only one initial mode, $k_0=1$, is linearly unstable, so that an initial perturbation with $k_0>1$ will decay exponentially back to base state (flat interface). On the other hand, the linearly unstable mode $k_0=1$ will first grow, then saturate to a traveling wave profile, and thus one fixed point can be identified in the $(\mathcal{E},\dot{\mathcal{E}})$ phase plane. Similarly, picking $k_m=2$ allows two linearly unstable modes, thus two fixed points in the energy phase plane. One fixed point is a period-one state, and the other is a period-two state.

However, while four unstable modes exists for $k_m=3$, only three fixed points are identified with periods two, three and four. When initialized with $k_0=1$, the period-one perturbation evolves and converges to a period-two traveling wave, as can be seen from Fig.~\ref{fig:energy plane 3D}. Note that $k_0=1$ is a special case in terms of the nonlinear interaction. For $k_0=1$, all normal modes in the system are harmonic components, so that $k=2,3,4$ will gain energy from $k_0=1$ also. A similar phenomenon can be observed in the $k_m=4$ slice of the energy phase plane. The initial perturbations with modes $k_0=2,3,4,5$ will evolve into states with corresponding $k_f=k_0$, while $k_0=1$ evolves into the period-three state.

Note that Fig.~\ref{fig:energy plane 3D} shows only four slices at integer $k_m$, but $k_m$ does not necessarily have to be an integer (because it is set by the non-integer system parameter $\alpha$ via Eq.~\eqref{eq:km}). 
Thus, our discussion only provides a representative view of the rich higher-dimensional dynamics. It is evident, from the four slices in Fig.~\ref{fig:energy plane 3D}, that bifurcations of fixed points occur as the parameter $k_m$ is varied. The number of fixed points increases with $k_m$, or more accurately, with the number of linearly unstable modes. Some fixed points move along the $\mathcal{E}$ axis with increasing $k_m$, such as the period-two and period-three states, while some disappear, like the period-one state. This observation partially answers the question of whether the number of traveling wave states will increase with $k_m$, and whether for certain states there is a possibly bounded ranged of $k_m$ allowing them. However, what exactly is this bound for each state, or each $k_m$, is beyond of the scope of this study. This question would be challenging, since as $k_m$ increases, more and more linearly unstable modes participate in the competition for setting the fundamental mode.

\subsection{Spectral stability of the traveling wave}
The tendency of a system to prefer a narrow set of states out of many possible ones is known as wavenumber selection \cite{ACDH86,QZC16}. In this section, we study this phenomenon by addressing the stability of these fixed points in the energy phase plane, focusing on the case of $k_m=4$.

To this end, we perturb the traveling wave profile, and numerically track the evolution of the perturbation via direct simulation of the PDE. We find that period-two and period-three states behave like local attractors, while period-four and period-five states are saddle points. We verify the type of the fixed points through spectral (in)stability analysis \cite{KP13,DKKSS14}. Specifically, we rewrite Eq.~\eqref{eq:e=delta2} in the moving frame with $\zeta=x-v_ft$, $\tau=t$ as:
\begin{equation}
    \eta_\tau-v_f\eta_\zeta = -\delta\alpha \eta_{\zeta\zeta} 
    + \beta \eta_{\zeta\zeta\zeta} - \eta_{\zeta\zeta\zeta\zeta} 
    + [(-\delta\alpha \eta_\zeta + \beta \eta_{\zeta\zeta} - \eta_{\zeta\zeta\zeta})\eta]_\zeta + \delta(\gamma \eta_\zeta^2 )_{\zeta\zeta},
\label{eq:linearization}
\end{equation}
with the propagation velocity $v_f$ calculated numerically. The perturbed traveling wave solution is written as $\eta(\zeta,\tau)=\Xi(\zeta)+\mathfrak{d}W(\zeta)e^{\mathcal{\lambda}\tau}$, where $\Xi(\zeta)$ is the stationary solution of Eq.~\eqref{eq:linearization} (hence, the traveling wave solution of Eq.~\eqref{eq:e=delta2}), and $\mathfrak{d}\ll1$ is an arbitrary perturbation parameter. Substituting the perturbed $\eta(\zeta,\tau)$ into Eq.~\eqref{eq:linearization}, and neglecting nonlinear terms, 
we obtain a linear eigenvalue problem
\begin{equation}
    \lambda W = \mathcal{L}W,\qquad \mathcal{L}:=\sum_{n=0}^{4} \mathcal{C}_n D_n.
\label{eq:linear eigenvalue}
\end{equation}
Here, the $\mathcal{C}_n=\mathcal{C}_n (D_0\Xi,\hdots,D_4\Xi,v_f)$ are vector-valued functions (see electronic supplementary material {\S}F for their expressions) of the traveling wave profile $\Xi(\zeta)$ and its gradients, and the differentiation matrices $D_n$ are discretizations of $\partial^n/\partial \zeta^n$ ($D_0=\bbm I$ is the $N\times N$ identity matrix) evaluated by the Fourier spectral approach \cite{STW11}. The eigenvalue problem in Eq.~\eqref{eq:linear eigenvalue} is solved numerically with \texttt{linalg.eig} from the NumPy stack in Python \cite{NumPy}. The spectrum was validated via a grid-independence study using grids with $N=256$, $512$, and $1024$ points.

Next, we use this numerical spectral stability approach to understand the state transitions and the stability of fixed points in the energy phase plane introduced in {\S}6\ref{sec:fp_epp}.

\subsection{The state transition process}
\begin{figure}[ht]
 \centering
 \includegraphics[keepaspectratio=true,width=0.99\columnwidth]{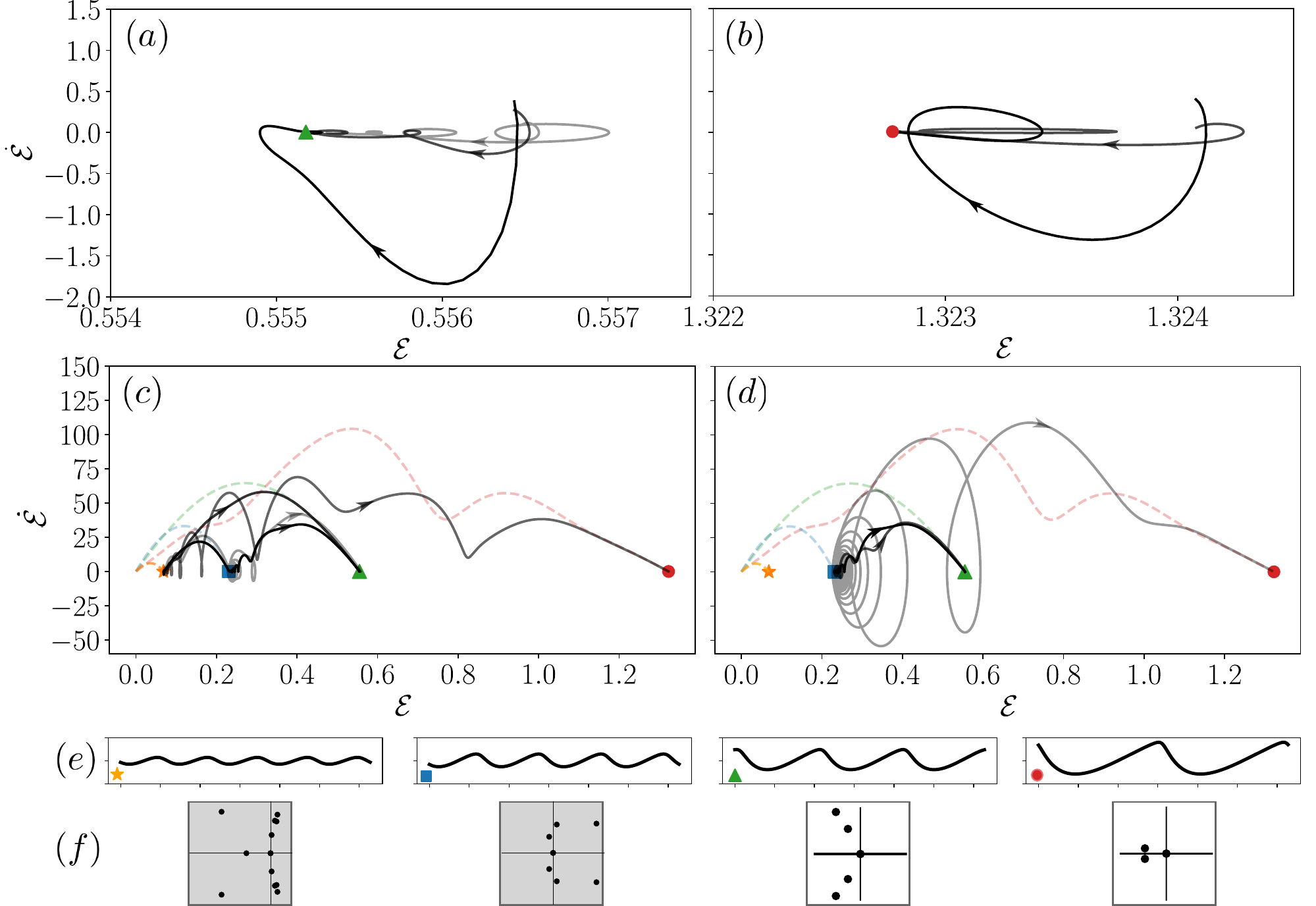}
 \caption{Stability diagram based on the energy phase plane. Perturbations around the attractors, corresponding to (a) the period-three and (b) the period-two traveling wave solutions, converge. State transitions are observed near the saddle points corresponding to  (c) the period-five and (d) the period-four traveling wave profiles.
 The solid curves' colors represent initial perturbations with different wavenumbers, which lead to different dynamics (and outcomes). The wave profiles are shown in (e), with the symbols in the corners of the plots denoting the corresponding fixed points in the phase planes in (a,b,c,d). In (f), the leading eigenvalues of the linearization about the corresponding wave profile in (e) are shown.}
 \label{fig:energpln_spect}
\end{figure}

Figure~\ref{fig:energpln_spect}(a,b) shows that the period-three and period-two fixed points, respectively, in the $(\mathcal{E},\dot{\mathcal{E}})$ phase plane are attractors. Small perturbations about them will decay, and the evolution will convergence back to the corresponding periodic traveling wave profiles. This observation can be confirmed by the spectral stability calculation, its results shown in Fig.~\ref{fig:energpln_spect}(f), which shows that all eigenvalues have negative real part, except for the zero eigenvalue, which represents the translational invariance of the traveling wave solution. 

On the other hand, Fig.~\ref{fig:energpln_spect}(c,d) show that the period-five and period-four fixed points, respectively, are saddles. A small perturbation around the period-four fixed point will grow and oscillate away, till the evolution converges to the period-three fixed point (an attractor), black curve in Fig.~\ref{fig:energpln_spect}(d). For a different perturbation, gray curve in Fig.~\ref{fig:energpln_spect}(d), this process can lead to convergence to the period-two attractor (see \texttt{output\_f4\_p2.mp4} in the electronic supplemental material for a video of this process). The perturbation evolution around the period-five fixed point, black curve Fig.~\ref{fig:energpln_spect}(c), is more interesting. It is featured by a two-stage transition process. First, the perturbation will first oscillate and grow rapidly, attracted to the neighborhood of the period-four fixed point. Then, it will oscillate away again, until finally converging to the period-three attractor (see \texttt{output\_f5\_p4.mp4} in the electronic supplemental material for a video of this process). These saddle point behaviors can be confirmed from the linear eigenspectra shown in Fig.~\ref{fig:energpln_spect}(f) as well. The period-four profile has two pairs of conjugate eigenvalues with positive real part, while the period-five profile has four pairs. 

\begin{figure}[t!]
 \centering
 \includegraphics[keepaspectratio=true,width=0.99\columnwidth]{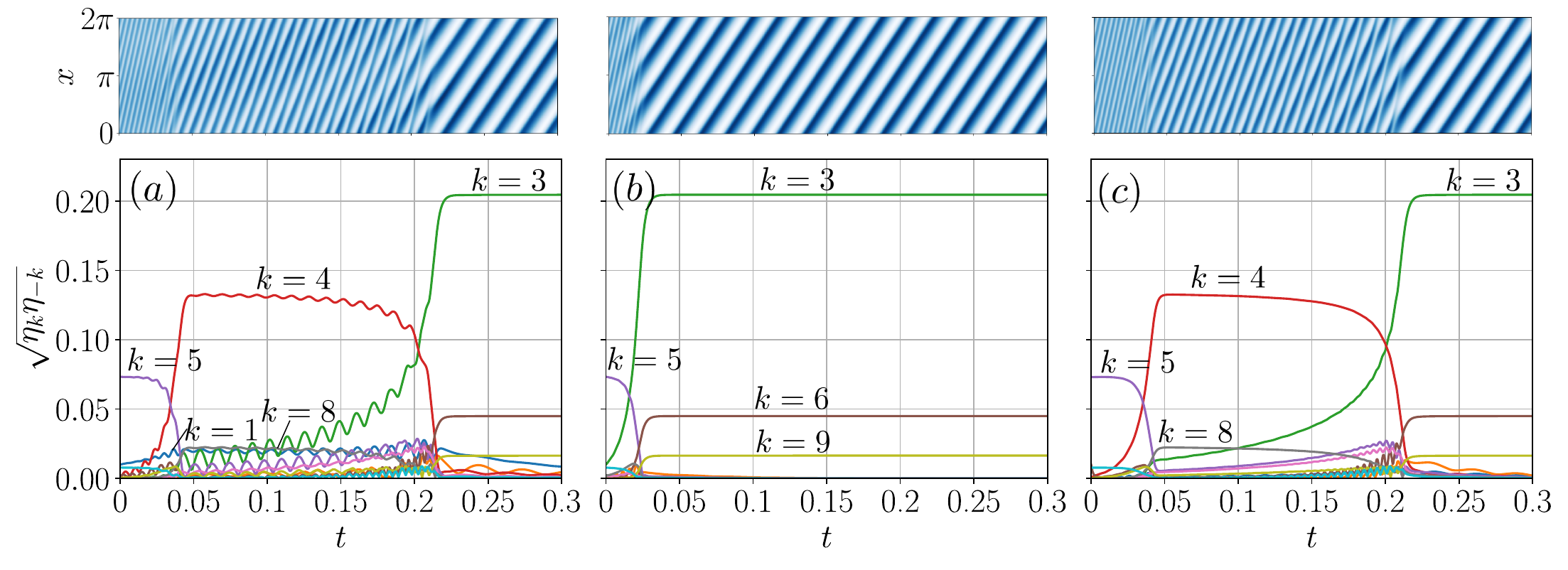}
 \caption{Fourier mode energy evolution (mode competition and nonlinear interaction) for a perturbed period-five traveling wave subjected to harmonic perturbation with (a) $k_p=1$, (b) $k_p=3$, and (c) $k_p=4$. The top row shows the corresponding space-time plot of the transition process with color representing the amplitude of the wave profile $\eta$.}
 \label{fig:energy transtion a5}
\end{figure}

A closer examination of the state transition process is shown in Fig.~\ref{fig:energy transtion a5} for three representative perturbations around the period-five fixed point. Rapid oscillation of the modes' energies can be observed during the transition process, indicating intense nonlinear interactions. The space-time plot shows a similar phase shift feature as seen during the collision of solitons \cite{ZK65}, but the wave profile is completely modified here. Figure~\ref{fig:energy transtion a5}(b) shows a one-stage transition due to a single-mode perturbation $\mathfrak{d}W(\zeta)=0.02\sin(k_p \zeta)$, $k_p=3$. This mode's energy $|\eta_3|$ increases exponentially, overtakes the initial $|\eta_5|$ value and converges to the period-three attractor. Figure~\ref{fig:energy transtion a5}(a,c) show a two-stage transition with single-mode perturbations $k_p=1,4$, respectively. Figure~\ref{fig:energy transtion a5}(a) shows higher level of oscillation than (c), since all modes are harmonics of $k_p=1$, and higher $|\eta_1|$ can be observed for $t\in[0.05,0.2]$. The interaction between mode 5 and mode 1 (Fig.~\ref{fig:energy transtion a5}(a)) immediately excites mode 4, and $|\eta_4|$ grows exponentially as the most unstable modes of the linear system. This results in a similar transition process for $k_p=1$ and $4$ in Fig.~\ref{fig:energy transtion a5}(a) and (c), respectively.

While such transition paths are complex and intriguing, we would like to emphasize the existence of the transition depends on the spectral stability of the traveling wave profile itself, which is interpreted as a saddle point or an attractor in the energy phase plane, and the transition direction is determined by the perturbation $W(\zeta)$. After an immediate targeted transition, whether another transition happens or not depends on the spectral stability of the subsequent wave profile attained.

Another intriguing aspect of this topic is multi-mode perturbations to the unperturbed flat interface, which is a more realistic situation that might arise in experiments, where the mode of the ambient noise is hard to control in an experiment. The competition between all possible states will finally select the observable pattern. Next we analyze this multi-mode case and provide an explanation of the selection process leading to multiperiodic nonlinear traveling waves.

\subsection{Multiperiodic waves}
\label{sec:multiperiod}

\begin{figure}[ht]
 \centering
 \includegraphics[keepaspectratio=true,width=\columnwidth]{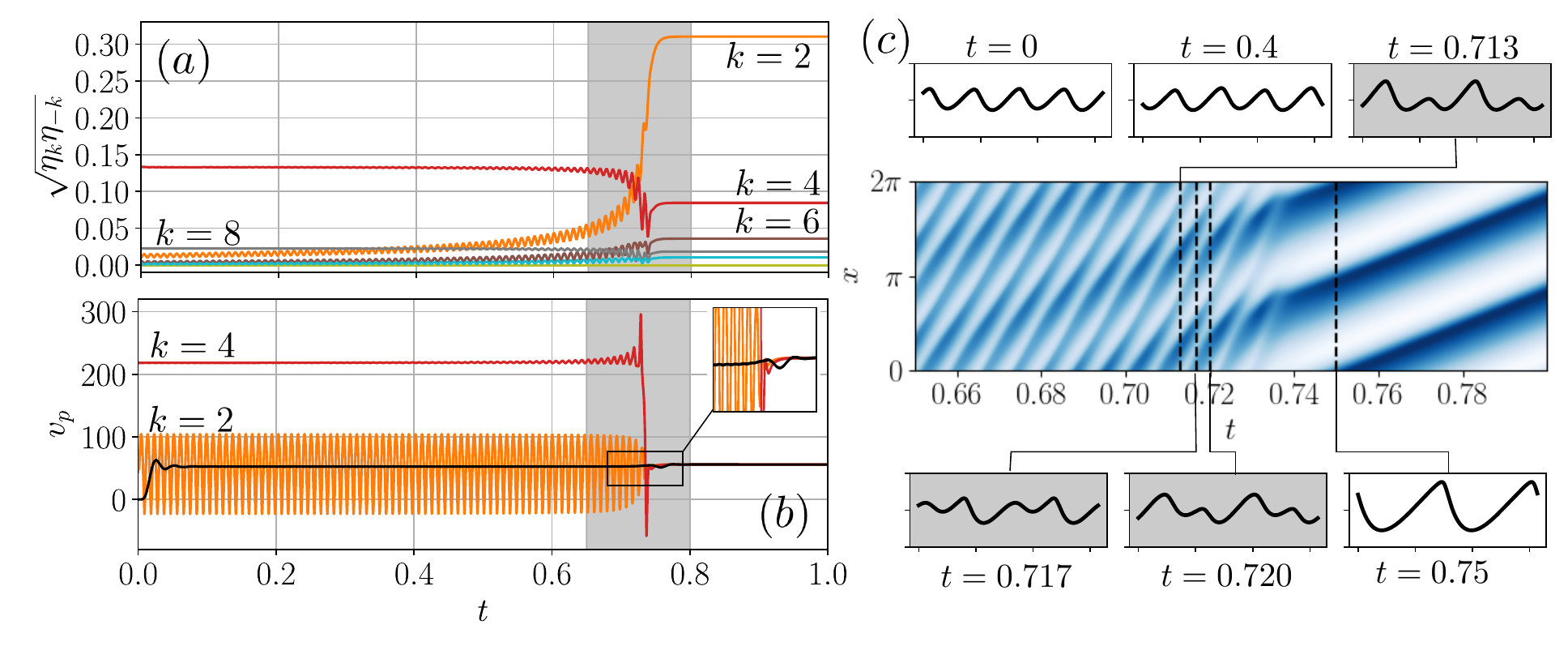}
 \caption{(a) Fourier modes energy interactions for a perturbed period-four traveling profile with perturbation $\mathfrak{d}W(\zeta)=0.02\sin(2 \zeta)$. (b) The phase velocities of mode 4 and mode 2. The black solid line shows the filtered $v_p(k=2)$. (c) Space-time plot and the corresponding wave profiles of the transition during $t\in[0.65,0.8]$, marked as the grey region in (a) and (b).}
 \label{fig:a2a4}
\end{figure}

An interesting observation from Fig.~\ref{fig:energy transtion a5}(a) is the coexistence of mode 1 and mode 4 during the transition, exemplified by the oscillations about the period-four fixed point in the energy phase plane shown in Fig.~\ref{fig:energpln_spect}(c). The energy components of the wave profile are harmonics of $k_f=4$, except the nontrivial $|\eta_1|\approx |\eta_8|$. During the time interval $t\in[0.05,0.2]$, the space-time plot of wave profile evolution  shows that a period-four wave is modulated by mode 1. This coexistence lasts for a relatively long time (compared to the total transition time) until mode 3 ultimately becomes dominant. An even longer coexistence is found when perturbing the period-four traveling profile with mode 2, as shown by the gray curve in Fig.~\ref{fig:energpln_spect}(d), leading to a period-four wave modulated by mode 2, as in Fig.~\ref{fig:a2a4}(a) (see \texttt{output\_f4\_p2.mp4} in the electronic supplemental material for a video of this process). The interaction between mode 2 and mode 4 occurs for $t\in[0,0.6]$, an interval twice longer than any complete transitions in Fig.~\ref{fig:energy transtion a5}. 

These long-lived multiperiodic waves states, which we have identified numerically, can be considered analogous to double cnoidal waves of the KdV equation. Double cnoidal waves are the spatially periodic generalization of the well-known two-soliton solution of KdV \cite{BH91}. They can be considered as exact solutions with two independent phase velocities \cite{HB91}. The evolution of the phase velocities $v_p(k)$ of modes $k=2$ and $4$ (of the Fourier decomposition of $\eta$) are shown in Fig.~\ref{fig:a2a4}(b). The phase velocity of mode 2 experiences more intense oscillations than mode 4, which can be seen also from Fig.~\ref{fig:a2a4}(a). These oscillations are caused by the energy interaction between even modes, and a low pass filter can be applied to evaluate a time-averaged phase velocity for mode 2, shown as the black curve (the jump around $t=0$ is a windowing effect). It is surprising to see that while $|\eta_2|$, the amplitude of mode 2, is growing slowly, its phase velocity maintains around $v_p(k=2)\approx 53.5$, which is independent of $v_p(k=4)\approx 218.1$. Haupt and Boyd \cite{HB91} constructed double cnoidal solutions of KdV through a harmonic balance of lower modes. On the other hand, the sharper peak of the quasi-double-cnoidal-waves in Fig.~\ref{fig:a2a4}(c) shows the importance of the balance among higher harmonic modes in our model KS-type long-wave equation.

The rapid transition during $t\in[0.65,0.8]$ is characterized by ``wave chasing" in the physical domain. Mode 2 and mode 4 become comparable in Fourier energy, with mode 4 propagating faster than mode 2. Visually, this observation is similar to soliton collisions: when the peak of mode 2 is caught by that of mode 4, an elevation can be observed and then a deperession as they separate, shown in the wave profiles in Fig.~\ref{fig:a2a4}(c) for $t\in[0.713,0.72]$. However, while soliton collision (in the sense of Zabusky and Kruskal \cite{ZK65}) leave the wave profile and propagation velocity unchanged, the ``chasing" (and interaction) in current study results in the waves ultimately merging into the  period-two nonlinear traveling wave. The phase velocity of mode 4 dramatically decreases, and all modes in the system merge into a phase velocity $v_p\approx 55.8$, which becomes the propagation velocity $v_f$ of the period-two traveling wave. 
It is interesting to note that, in this process, the time-averaged propagation velocity of mode 2 barely changes, except for the mitigation of the oscillations. This can be intuitively understood from the strong stability of the period-two traveling wave profile, while a mathematical reason might emerge from the singular limit of a double cnoidal wave (if it exists in this system).

In the end, this study answers one question posed in \cite{YC21}:  when the energy of higher modes is dominant, this confined ferrofluid system can accommodate multiperiodic traveling waves, resembling a long-lasting, but non-integrable, double cnoidal wave field. When the energy of the two component modes becomes comparable, a rapid transition happens and the modulated propagating wave profile saturates to its envelope. 
In a sense, this means that these periodic nonlinear waves lose their shapes upon ``collision." However, it would be interesting to ask if a localized solitary wave also exists for our model equation, and to address what would happen during the localized waves' collisions.


\section{Conclusion}
\label{sec:conclusion}

The dynamics of long, small-amplitude nonlinear waves on the interface of a thin ferrofluid film was analyzed for the configuration of a horizontal Hele-Shaw flow subjected to a tilted magnetic field. We showed that such ferrofluid interfaces support periodic traveling waves governed by a modified KS-type equation, which we derived. A linear stability analysis and a nonlinear energy budget were employed to reveal that the balance between stabilizing surface tension forces (energy sink/loss) and destabilizing magnetic forces (energy source/gain) leads to the generation of dissipative solitons on the ferrofluid interface. The effect of key parameters was investigated, and the corresponding magnetic field configurations were discussed. Our results lead to quantitative understanding of these nonlinear periodic traveling wave profiles, and how interfacial waves can be generated and controlled (specifically, their propagation velocity and shape) non-invasively by an external magnetic field. A multiple-scale analysis provides the correction a weakly nonlinear correction to the propagation velocity of harmonic waves. This calculation also reveals how the marginally unstable linear solution is equilibrated by the weak nonlinearity and tends to the permanent traveling wave solution. At the same time, the model equation~\eqref{eq:e=delta2} features a variety of interesting novel nonlinearities that could open avenues of future mathematical research.

In this respect, we identified the allowed wave states (specifically, their spatial periods), which bifurcate as the most unstable linear mode $k_m$ is varied, as fixed points in an energy phase plane, using the  dissipative soliton concept \cite{CV95}. State transitions are observed when some traveling wave profiles are perturbed, depending on their spectral stability, and the transition ``direction'' (towards another fixed point in the energy phase plane) is determined by the perturbation. It would be of interest to realize the obtained traveling wave profiles (and their transition dynamics) in laboratory experiments. The wave selection process with multi-mode perturbations poses a challenge in that the initial perturbation must be carefully controlled, especially for the spectrally unstable profiles.

Another novel feature of this study is that multiperiodic nonlinear waves (akin to the double cnoidal wave of the KdV equation) were found numerically in the context of a (non-integrable) long-wave equation of the modified KS type. Perturbations of spectrally stable modes interact intensely with their harmonics, which are already present as part of the original spectrally unstable traveling wave profile. Such interactions are long-lived, until an abrupt transition to a final stable traveling wave occurs. As mentioned in {\S}6\ref{sec:multiperiod}, we were unable to construct perturbative solutions in the sense of the double cnoidal waves \cite{HB91}, therefore a complete mathematical explanation of these multiperiodic nonlinear wave dynamics (and the transitions between them) remains an open problem to be addressed in future work. Finally, it would also be of interest to derive a 2D version of our model long-wave equation, and the dynamics it governs could be compared and contrasted to recent work on the 2D KS equation \cite{KKP15}.



\dataccess{Python script required to run and analyze the numerical simulations of Eq.~\eqref{eq:e=delta2}, and resulting data files are available at: \url{https://github.com/zongxin/long_wave_eq_ferrofluid_thin_film}.}

\aucontribute{Both authors contributed equally to the analysis of the problem and the derivation of the mathematical model. I.C.C.\ initiated the project and supervised the analysis. Z.Y.\ conducted all the case studies, simulations, and data analysis. Both authors discussed the results and contributed equally to the final version of the manuscript.}

\competing{We declare that we have no competing interests.}

\funding{This research was supported by the U.S.\ National Science Foundation under grant no.~CMMI-2029540 (to I.C.C.) and a Ross Fellowship from The Graduate School at Purdue University (to Z.Y.).}

\ack{I.C.C.\ thanks G.M.\ Homsy for providing a copy of reference \cite{H74} and comments on interfacial waves.}


\bibliographystyle{rspublicnatwithsort.bst}
\setlength{\bibsep}{0pt}
\renewcommand{\bibfont}{\small}
\bibliography{Mendeley_refs,yu_refs}


\clearpage

\appendix

\setcounter{page}{1}
\renewcommand{\theequation}{\thesection.\arabic{equation}}

\begin{center}
\color{jobcolor}
{\LARGE
\textsf{Supplementary Appendices for}\\
\textsf{``Long-wave equation for a confined ferrofluid interface: Periodic interfacial waves as dissipative solitons''}}

\bigskip
\textsf{\Large by Zongxin Yu and Ivan C.\ Christov}
\end{center}

\section{The constants \texorpdfstring{$B_n$}{Bn}}
\label{apdx:ABCD}
The expressions for the constants in Eq.~\eqref{eq:boundary with ABCD} in the main text are:
\begin{subequations}\begin{align}
B_1 &= 2\varepsilon[-\nby(1+\chi) +\nbx]-2\varepsilon^2[(1+\chi)\nby+3\nbx],\\
B_2 &= 2\chi \sqrt{\nbx\nby},\\
B_3 &=\chi [\nby(1+2\varepsilon)-\nbx(1-2\varepsilon)+(\nby-3\nbx)\varepsilon^2],\\
B_4 &= \varepsilon^2[(1+\chi)\nby+3\nbx].
\end{align}\end{subequations}

\section{The long wave equation with \texorpdfstring{$\varepsilon=\mathcal{O}(\delta)$}{vareps=O(delta)}}
\label{apdx:e=delta}

Setting $\varepsilon=\delta$ and performing the same re-scaling as given in Eq.~\eqref{eq:long time scal} in the main text, the interface evolution Eq.~\eqref{eq: general eta} for $\varepsilon=\mathcal{O}(\delta)$ reads:
\begin{equation}
\eta_t=(-\alpha-\delta\vartheta) \eta_{xx} +\beta \eta_{xxx}  -\eta_{xxxx}
+\{[(-\alpha-2\delta\vartheta) \eta_{x} +\beta \eta _{xx}  -\eta_{xxx} ]\eta\}_x
+ \delta\gamma (\eta_x^2)_{xx} .
\label{eq:e=delta}
\end{equation}
Observe that, whether $\varepsilon=\mathcal{O}(\delta)$ or $\varepsilon=\mathcal{O}(\delta^2)$, the resulting nonlinear evolution equation has a similar structure, since terms multiplied by $\delta \vartheta$ are small in comparison with the dominant $\alpha$ terms in Eq.~\eqref{eq:e=delta} ($\alpha\approx\vartheta$ in the small-$\nbx$ regime within the scope of this study). 

The main difference between the two scalings is the magnitude of individual terms, \textit{e.g.}, terms multiplied by $\alpha$ in Eq.~\eqref{eq:e=delta} can be compared to those multiplied by $\delta\alpha$ in Eq.~\eqref{eq:e=delta2}. 
In this study, we are interested in traveling wave solutions, so that the coefficients $\alpha$ in Eq.~\eqref{eq:e=delta} (and $\delta\alpha$ in Eq.~\eqref{eq:e=delta2}) are kept within a certain range. Therefore, since $\alpha$, $\beta$ and $\gamma$ are expressed in terms of $\nbx$ and $\nby$ (via Eq.~\eqref{eq: main parameters}), then the different ranges for $\alpha$ in Eq.~\eqref{eq:e=delta} and Eq.~\eqref{eq:e=delta2} necessarily leads to different ranges for $\beta$ and $\gamma$ in these equations (for given $\nbx$ and $\nby$).

For example, in Eq.~\eqref{eq:e=delta2}, the periodic wave with $k_m=4$ requires $\alpha=320$ in Eq.~\eqref{eq:e=delta2} and $\alpha=32$ in Eq.~\eqref{eq:e=delta}. If we stay within the small tilt angle assumption, \textit{i.e.}, $q=0.01$ according to Eq.~\eqref{eq:parameters rho q }, then to maintain similar stability, the system where $\varepsilon=\mathcal{O}(\delta^2)$ requires a stronger magnetic field, \textit{i.e.}, $\rho|_{\varepsilon=\mathcal{O}(\delta^2)}=(1/\delta) \rho|_{\varepsilon=\mathcal{O}(\delta)}$. Correspondingly, $\beta|_{\varepsilon=\mathcal{O}(\delta^2)}\approx 16.1$, while $\beta|_{\varepsilon=\mathcal{O}(\delta)}\approx 16.1\delta$. That means Eq.~\eqref{eq:e=delta} is subjected to weaker dispersion, if we require that the base states under Eq.~\eqref{eq:e=delta2} and Eq.~\eqref{eq:e=delta} both have the same linear stability properties.

\section{Grid convergence and time step refinement}
\label{apdx:gird convg}

\begin{figure}[ht]
 \centering
 \includegraphics[keepaspectratio=true,width=\columnwidth]{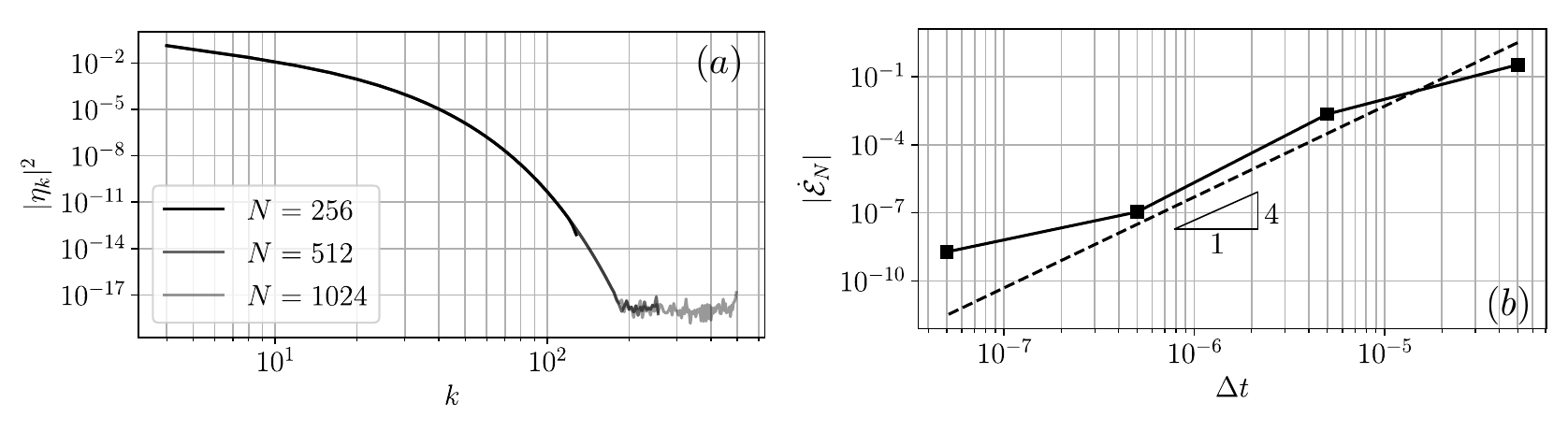}
 \caption{(a) Spectral energy of harmonic modes ($k = k_0, 2k_0, 3k_0,\hdots$, connected as a curve to ``guide the eye'') for the period-four traveling wave with $k_0=4$ at $k_m=4$, $q=0.01$, and $\Delta t = 5\times10^{-8}$. (b) The numerically computed dissipation rate error $\dot{\mathcal{E}}_N$ decreases with time step refinement for the same physical parameters as (a) and $N=512$.}
 \label{fig:grid convergence}
\end{figure}

A grid convergence study with three levels of the grid resolution was conducted to validate the pseudospectral method with ETDRK4 time stepping introduced in {\S}4\ref{sec:numerical_method} for our model PDE~\eqref{eq:e=delta2}. In this appendix, we demonstrate the grid convergence for the period-period traveling wave at $k_m=4$, which is the most frequently discussed case in the main text. Figure~\ref{fig:grid convergence}(a) shows that the energy of harmonic modes decays with the wavenumber, and ``piling up" occurs near the ``tail'' on the grids with $N=512$ and $N=1024$. This phenomenon is due to hitting the limit of double precision floating point arithmetic, which is indicative of spectral convergence. Additionally, the results on the grid with $N=512$ match well with those of $N=1024$. Actually, the grid with $N=256$ also provides a satisfactory result for the large scales ($k\in[4,128]$) but the grid with $N=512$ can resolve smaller scales better. Therefore, Fig.~\ref{fig:grid convergence}(a) supports our decision to use $N=512$ for our simulations in the main text.

When the traveling wave solution is obtained, the energy change rate is supposed to reach a steady state, \textit{i.e.}, $\dot{\mathcal{E}}=0$, and the right-hand side of Eq.~\eqref{eq:energy budget} vanishes as well. However, due to the numerical truncation, the sum of all the energy production/dissipation terms on the right-hand side actually changes with the time step $\Delta t$. Specifically, denote as $\dot{\mathcal{E}}_N=
\dot{\mathcal{E}}-
\int_0^{2\pi} 
 \delta\alpha \eta_x^2
- \eta_{xx}^2
+\delta\alpha \eta_{x}^2\eta
+\frac{1}{2}\beta\eta_{x}^3 
-\eta \eta_{xx}^2
\,dx
$ the error in the numerically computed dissipation rate. 
Four time steps $\Delta t=5\times10^{-5}$, $5\times10^{-6}$, $5\times10^{-7}$, $5\times10^{-8}$ were considered for verification, and all give qualitatively consistent results. It is noteworthy that the time step spans two orders of magnitude, while the ETDRK4 scheme is still stable for this fourth-order stiff PDE. The dissipation rate error $\dot{\mathcal{E}}_N$ is shown in Fig.~\ref{fig:grid convergence}(b), and it exhibits fourth-order convergence with respect to $\Delta t$. Since time-step-convergence has been demonstrated, for this study (and the results in the main text), we use the intermediate time step size $\Delta t=5\times10^{-8}$, which commits a dissipation rate error of $\dot{\mathcal{E}}_N=1.91\times10^{-9}$, as a compromise between numerical accuracy and computational cost.

\section{Propagation velocity calculation from simulation} \label{apdx:propagation velocity}

\begin{figure}[t]
 \centering
 \includegraphics[keepaspectratio=true,width=\columnwidth]{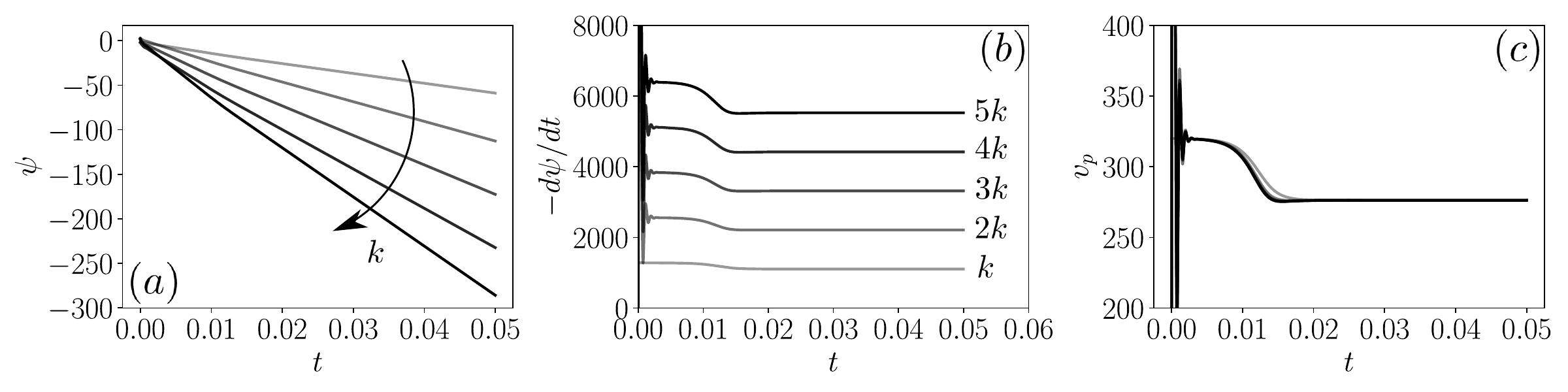}
 \caption{(a) Evolution of the modes' phases computed via a Fourier transform. (b) The time-rate-of-change of the phases. (c) Evolution of the phase velocities of the first five harmonic modes.} 
 \label{fig:cal vf}
\end{figure}

The Fourier modes comprising the nonlinear traveling wave profile are given by $\eta_{nk}(t)=c_ne^{-in\omega(k) t}$, with constant $c_n\in\mathbb{C}$ that account for their relative phases. The phase $\psi(t;k)=\angle \eta_{nk}=\angle c_n-n\omega(k) t$ can be computed trough a Fourier transform, as shown in Fig.~\ref{fig:cal vf}(a). Its rate of change, $-d\psi/dt=n\omega(k)$, is shown in Fig.~\ref{fig:cal vf}(b). The phase velocity $v_p=n\omega(k)/nk$ becomes independent of $k$ when the permanent traveling wave solution is attained upon nonlinear saturation of the unstable small-perturbation initial condition. In other words, all modes propagate at the same velocity, as shown in Fig.~\ref{fig:cal vf}(c), in the final state. The mean final phase velocities of first five harmonics are used to evaluate the propagation velocity as  $v_f=\frac{1}{5}\sum_{n=1}^{5} v_p(nk)$. Note that this approach can only be applied for the permanent traveling wave solution; the initial transition time period in Fig.~\ref{fig:cal vf} (before the permanent profile is attained) is a meaningless transient.

\section{Propagation velocity via multiple-scale analysis} \label{apdx:multi-scale}
In this appendix, we perform a multiple-scale analysis of harmonic wave propagation, following the approach outlined by Kevorkian and Cole \cite[Ch.~6]{KC96}.
Introducing the slow time and traveling wave coordinate $\xi=kx-\omega_p t$ as in the main text, we obtain the following transformations of partial derivatives: 
\begin{equation}
\p_t = -\omega_p \p_\xi +\mathfrak{e}^2 \p_{t_2},\qquad
\p_x=\p_\xi.
\label{eq:MS_deriv_transform}
\end{equation}
Next, substituting the derivative transformations from Eq.~\eqref{eq:MS_deriv_transform} and the dependent variable expansion from Eq.~\eqref{eq: y expansion} into the weakly nonlinear equation~\eqref{eq:weak nonlinear} gives rise to a series of problems at each order of $\mathfrak{e}\ll1$.

\subsection*{$\bullet\quad \mathcal{O}(1):$}
Denoting the linear operator as $\mathbb{L}$, the leading-order equation is given as:
\begin{equation}
\mathbb{L}[\mathcal{Y}_0]=
\left(-\omega_p \p_\xi 
+k_f^2 k^2 \p_\xi^2 
-\beta k^3\p_\xi^3 
+k^4\p_\xi^4 \right)\mathcal{Y}_{0}=0.
\label{eq:y0}
\end{equation}
In this study, we are interested in the phase velocity of a single harmonic wave with wavenumber $k$, which is also subjected to weak linear instability if $\varkappa>0$. So, we set $k_f=k$ and $\omega_p=k^3\beta$. Then, the general solution of Eq.~\eqref{eq:y0} is 
\begin{equation}
    \mathcal{Y}_0(\xi,t_2) = A_0(t_2)e^{i\xi} + c.c.,
\label{eq:y0 general}
\end{equation}
where $c.c.$ stands for complex conjugate.

\subsection*{$\bullet\quad \mathcal{O}(\mathfrak{e}):$}
At this order, we obtain an inhomogeneous PDE:
\begin{equation}
\mathbb{L}[\mathcal{Y}_1]
= \left[(-\delta\alpha \mathcal{Y}_{0,\xi}
+\beta \mathcal{Y}_{0,\xi\xi}  
- \mathcal{Y}_{0,\xi\xi\xi} )\mathcal{Y}_0\right]_{\xi} 
+ \delta(\gamma \mathcal{Y}_{0,\xi}^2 )_{\xi\xi}.
\label{eq:y1}
\end{equation}
Substituting Eq.~\eqref{eq:y0 general} into Eq.~\eqref{eq:y1} we have:
\begin{equation}
\mathbb{L}[\mathcal{Y}_1] 
= 2[\delta\alpha k^2-i\beta k^3+(2\delta\gamma-1)k^4]A_0^2e^{2i\xi}+c.c.
\label{eq:y1 with y0}
\end{equation}
The general solution of this PDE, denoted $\mathcal{Y}_1$, can be written as
\begin{equation}
\mathcal{Y}_1 (\xi,t_2) = A_1(t_2) e^{i\xi} + A_{1p}(t_2)e^{i2\xi} + c.c.,
\label{eq:y1 general}
\end{equation}
where $\mathcal{Y}_{1p}=A_{1p}e^{i2\xi}+c.c.$ is the particular solution. Substituting $\mathcal{Y}_{1p}$ into Eq.~\eqref{eq:y1 with y0}, we obtain
\begin{equation}
A_{1p}=p A_0^2, \quad \text{with}\quad p=\frac{[\delta\alpha -i\beta k+(2\delta\gamma-1)k^2]}
{6k^2+3i\beta k}.
\end{equation}

\subsection*{$\bullet\quad \mathcal{O}(\mathfrak{e^2}):$}
At this order we obtain
\begin{equation}
\begin{aligned}
\mathbb{L}[\mathcal{Y}_2]
=-&( 
\mathcal{Y}_{0,t_2}+ \varkappa k^2 \mathcal{Y}_{0,\xi\xi}
)\\
-&\delta\alpha k^2 (
\mathcal{Y}_{1,\xi\xi}\mathcal{Y}_0 
+\mathcal{Y}_{0,\xi\xi}\mathcal{Y}_1
+2\mathcal{Y}_{1,\xi}\mathcal{Y}_{0,\xi}
)\\
+&\beta k^3 (
\mathcal{Y}_{1,\xi\xi\xi}\mathcal{Y}_0 
+\mathcal{Y}_{0,\xi\xi\xi}\mathcal{Y}_1
+\mathcal{Y}_{1,\xi\xi}\mathcal{Y}_{0,\xi}
+\mathcal{Y}_{0,\xi\xi}\mathcal{Y}_{1,\xi}
)\\
-&k^4 (
\mathcal{Y}_{1,\xi\xi\xi\xi}\mathcal{Y}_0 
+\mathcal{Y}_{0,\xi\xi\xi\xi}\mathcal{Y}_1
+\mathcal{Y}_{1,\xi\xi\xi}\mathcal{Y}_{0,\xi}
+\mathcal{Y}_{0,\xi\xi\xi}\mathcal{Y}_{1,\xi}
)\\
+&2\delta\gamma k^4(
\mathcal{Y}_{1,\xi\xi\xi}\mathcal{Y}_{0,\xi}
+\mathcal{Y}_{0,\xi\xi\xi}\mathcal{Y}_{1,\xi} 
+2 \mathcal{Y}_{1,\xi\xi}\mathcal{Y}_{0,\xi\xi}
)
\end{aligned}
\label{eq:y2}
\end{equation}
Substituting the previously obtained solutions $\mathcal{Y}_0$ and $\mathcal{Y}_1$ into Eq.~\eqref{eq:y2}, we have 
\begin{equation}
    \begin{aligned}
        \mathbb{L}[\mathcal{Y}_2]=
        &-( A_{0,t_2}-\varkappa k^2A_{0})e^{i\xi}\\
        &+d_0 A_{1}A_0^*
        +d_1A_{1p}A_0^*e^{i\xi}
        +d_2A_{1}A_0e^{2i\xi}
        +d_3A_{1p}A_0e^{3i\xi}+c.c,
    \end{aligned}
    \label{eq:y2_2}
\end{equation}
where $d_i$ are complex constant coefficients. We are only concerned with  $d_1=[\delta\alpha k^2-5i\beta k^3-(7+4\delta\gamma)k^4]$. 

To eliminate the secular term in Eq.~\eqref{eq:y2_2}, we require that
\begin{equation}
    -( A_{0,t_2}-\varkappa k^2A_{0}) +d_1A_{1p}A_0^*=0,
\end{equation}
which gives rise to the amplitude equation
\begin{equation}
    A_{0,t_2} = \varkappa k^2A_{0} -Q|A_0|^2 A_0,
\label{eq:amplitude}
\end{equation}
which is known as the Landau equation, with $Q=-d_1p$. Let $A_0=\mathfrak{a}e^{i\mathfrak{b}}$, where $\mathfrak{a}$ and $\mathfrak{b}$ are real numbers. Then, the balance of the real and imaginary parts of Eq.~\eqref{eq:amplitude} gives:
\begin{align}
    \frac{d\mathfrak{a}}{dt_2} &= \varkappa k^2 \mathfrak{a}-\Re{[Q]}\mathfrak{a}^3,\\
    \frac{d\mathfrak{b}}{dt_2} &= -\Im{[Q]}\mathfrak{a}^2.
\end{align}
If $\Re{[Q]}>0$, which is true after substituting the simulation parameters, three fixed point can be identified, with $\mathfrak{a}=0$ being an unstable equilibrium point and $\mathfrak{a}=\pm \sqrt{\varkappa k^2/\Re{[Q]}}$ being stable. The long-time behavior, as $t_2\rightarrow \infty$, is that $\mathfrak{a}$ converges to these equilibrium points, and 
\begin{equation}
    \mathfrak{b}(t_2)
     \sim -\Im{[Q]}\frac{\varkappa k^2}{\Re{[Q]}}t_2+\mathfrak{b}_0 \quad\text{ as }\quad {t_2\rightarrow \infty}.
\end{equation}
Recall that $\mathcal{Y}_0(\xi,t_2)=A_0(t_2)e^{i\xi}+c.c.=\mathfrak{a}(t_2)e^{i(\xi+\mathfrak{b}(t_2))}+c.c.=2\mathfrak{a}(t_2)\cos\big(\xi+\mathfrak{b}(t_2)\big)$, with $t_2=\mathfrak{e}^2 t$, then the solution at the leading order can be obtained as Eq.~\eqref{eq:y_0 sol}.

\section{The functions \texorpdfstring{$\mathcal{C}_n$}{Cn} in the linear operator} 
\label{apdx:linear operator}

The functions arising in the linear operator $\mathcal{L}$ in Eq.~\eqref{eq:linear eigenvalue} in the main text are:
\begin{equation}
\begin{alignedat}{3}
\mathcal{C}_0 &=\delta \alpha \Xi_{\zeta\zeta} + \beta \Xi_{\zeta\zeta\zeta} - \Xi_{\zeta\zeta\zeta\zeta},\quad 
&\mathcal{C}_1 &=v_f+2\delta \alpha \Xi_{\zeta}+ \beta \Xi_{\zeta\zeta} -\Xi_{\zeta\zeta\zeta}+2\delta\gamma, \quad && \\
\mathcal{C}_2 &=\delta \alpha(1+\Xi)+ \beta\Xi_{\zeta} +4\delta\gamma \Xi_{\zeta\zeta},\quad
&\mathcal{C}_3 &= \beta(1+\Xi)- \Xi_{\zeta} +2\delta\gamma \Xi_{\zeta},\quad
&\mathcal{C}_4 &= -(1+\Xi).
\end{alignedat}
\end{equation}

\end{document}